\documentclass[prc,aps,showpacs,showkeys,groupedaddress,floatfix,superscriptaddress]{revtex4-1}
\usepackage{epsfig}
\usepackage{bm}
\usepackage{graphics}
\usepackage{graphicx}
\usepackage{epsfig}
\usepackage{amssymb}
\usepackage{amsmath}
\usepackage{color}
\usepackage{natbib}
\usepackage{amssymb}
\usepackage{amsmath}
\usepackage{graphicx}
\usepackage{subcaption}
\usepackage{graphicx}
\usepackage{rotating}
\usepackage{color}

\begin{document}

\title{The generalized Klein–Gordon oscillator in a Cosmic Space-Time with a Space-Like Dislocation and the Aharonov-Bohm Effect}

\author{B. C. L\"{u}tf\"{u}o\u{g}lu}
\email{Corresponding author : bclutfuoglu@akdeniz.edu.tr}

\affiliation{Department of Physics, Akdeniz University, Campus 07058 Antalya, Turkey}
\affiliation{Department of Physics,  University of Hradec Kr\'{a}lov\'{e}, Rokitansk\'{e}ho 62, 500\,03 Hradec Kr\'{a}lov\'{e}, Czechia}

\author{J. Kříž}
\affiliation{Department of Physics,  University of Hradec Kr\'{a}lov\'{e}, Rokitansk\'{e}ho 62, 500\,03 Hradec Kr\'{a}lov\'{e}, Czechia}

\author{P. Sedaghatnia}
\affiliation{Faculty of Physics, Shahrood University of Technology, Shahrood, Iran  P. O. Box : 3619995161-316.}

\author{H.  Hassanabadi}
\affiliation{Department of Physics,  University of Hradec Kr\'{a}lov\'{e}, Rokitansk\'{e}ho 62, 500\,03 Hradec Kr\'{a}lov\'{e}, Czechia}
\affiliation{Faculty of Physics, Shahrood University of Technology, Shahrood, Iran  P. O. Box : 3619995161-316.}

\date{\today}

\begin{abstract}
In the present work, we investigated the quantum behavior of a charged particle that is under the effect of a uniform magnetic external field. We assumed that space-time has a space-like dislocation with an internal magnetic flux. We examined two different types of potential energies that are known as the pseudo harmonic and Cornell type potential energies, with the nonminimal coupling.   We observed the Aharonov-Bohm effect in both cases. We extended the analysis on the energy spectrum functions to the different limits.
\end{abstract}
\maketitle
	

\section{Introduction}

Klein-Gordon (K-G) oscillator, which was first predicted by Bruce and Minning in 1993 \cite{1, 1b}, has been used in many studies since then. For example, Ahmed recently employed the K-G oscillator to investigate the effect of topological defects on the interactions of charged particles in the uniform magnetic field \cite{2}. A short while ago, Vit\'oria and Bakke introduced a space-time with a space-like dislocation to examine a charged particle's dynamic within a uniform magnetic field \cite{3}. We noticed that they employed the same metric with the same deformation to investigate the solutions of K-G oscillators that have position-dependent mass-energy with linear \cite{4} and Coulomb-type \cite{5}  in their previous studies as well.  K-G oscillator was also investigated in Minkowski space-time by using Coulomb-type potential energy with different coupling \cite{6, 7}. Boumali and Messai considered a cosmic string background and investigated the K-G oscillator with and without magnetic field \cite{8}. Hossseini \emph{et al.} revisited the problem with a Cornell potential \cite{9}. Note that, the Cornell potential energy was explored in Minkowski space-time as well \cite{10}. Carvalho \emph{et al.} employed a linear potential energy  to examine a K-G oscillator in the background of a space-time deduced from topological defects via the Kaluza-Klein theory \cite{11}. In very recent work, Ahmed revisited the same problem with a generalized K-G oscillator \cite{12}. In addition to these studies, the K-G oscillator was investigated in other space-times such as G\"odel-type \cite{13}, Som-Raychaudhuri type \cite{14}, etc.

In 1959, Aharonov and Bohm predicted a pure quantum effect \cite{15}, which would later be referred to by their own name \cite{16}. According to this effect, a phase shift in wave function will occur due to a magnetic flux that is restricted to where the particle cannot reach conventionally. Such a non-local effect is explained by the use of a gauge-dependent vector potential, which extends to the whole space beyond the region with a finite magnetic field.  Besides, it is related to the topological properties of a non-trivial background of the wave function \cite{17}.

The exact solution of a quantum system provides the necessary information explicitly. However, exactly solvable systems are very rare in practice. Therefore, alternative solution methods such as quasi-exact solution (QES), Nikiforov-Uvarov (NU) and etc. are being used. The QES method is an algebraic technique that determines a finite number of eigenstates and eigenvalues, which is the weak point of the method, with some ad hoc couplings \cite{18, Ushveridze}.  The NU method is another algebraic technique which is based on solving the linear differential equations in the second-order \cite{19}.  However, the weak point of the method is that every differential equation cannot be converted into the required differential equation of the NU method.  Despite all of this, we observe that both\ methods are frequently being employed in recent studies \cite{20, Znojil16, Quesne17, Quesne18, Znojil94, Znojil06, Bulent, Cari}. However, we would like to mention that the NU method can deduce spurious results as well \cite{Orhan}.

In this work, our main motivation is to investigate a generalized K-G oscillator that is in interaction with an external field including an internal magnetic flux field in a space-time with a space-like dislocation in the presence of Cornell and Pseudo harmonic potentials. To explore the AB effect we employ either the NU or the QES algebraic methods in the determination of the energy eigenvalues.

The rest of this paper is organized as follows: In section \ref{sec2} we derived the generalized K-G equation with a nonminimal coupling of the external magnetic field in the cosmic string space-time with a space-like dislocation with an internal magnetic flux.  In section \ref{sec3}, at first, we attempted to obtain a solution of the relativistic energy levels of the generalized K-G oscillator in the presence of Pseudo harmonic Potential by the NU method. Then we carried a discussion on the energy spectrum within physically meaningful limits. In section \ref{sec4} we employed the QES method to derive the energy eigenvalue function in the presence of Cornell-type potential energy. Alike the discussion that has carried on in section \ref{sec3}, we analyzed the energy spectrum function in various limits. Moreover, in both sections, we discussed the AB effect and degeneracy of the eigenvalues. After all, we presented a brief conclusion in section \ref{sec5}.

\section{The K-G oscillator  in a space-like dislocation}\label{sec2}	

We consider a metric in cylindrical coordinates  with the following line element \cite{H1, H2, H3, H4, HMO, HHK, HH3, HKC, HH2}
\begin{eqnarray}\label{1}
ds^{2}=-dt^{2}+dr^{2}+\alpha^{2}r^{2}d\varphi^{2}+(dz+\chi d\varphi )^{2},
\end{eqnarray}
where the variables $(r, \varphi, z)$ varies in the interval as follows:  $r\geq 0 $, $0\leq \varphi \leq 2\pi $ and $-\infty<z<\infty$.  Here, $\alpha$ is the angular parameter and it is proportional  to the linear mass density, $\mu$, via the relation $\alpha=1-4\mu$. We denote the spatial dislocation with the parameter $\chi$ \cite{HMO, HHK}. We express the components of the metric and its inverse matrix as
\begin{align}\label{2}
g_{\zeta \xi}=\left( \begin{matrix}
-1
&0
&0
&0
\\
0
&1
&0
&0
\\
0
&0
&\alpha^{2} r^{2}+\chi^{2}
&\chi
\\
0
&0
&\chi
&1
\end{matrix}\right)\qquad ,\qquad
g^{\zeta \xi}=\left( \begin{matrix}
-1
&0
&0
&0
\\
0
&1
&0
&0
\\
0
&0
&\frac{1}{\alpha^{2} r^{2}}
&\frac{-\chi}{\alpha^{2}r^{2}}
\\
0
&0
&\frac{-\chi}{\alpha^{2}r^{2}}
&1+\frac{\chi^{2}}{\alpha^{2}r^{2}}
\end{matrix}\right).
\end{align}
In this paper we examine the dynamics of relativistic quantum particles in a curved space-time. Therefore, we use the K–G equation that is given by \cite{5, 9, H4}
\begin{eqnarray}\label{3}
\left[\frac{1}{\sqrt{-g}}D_{\zeta}\big(\sqrt{-g} g^{\zeta \xi}D_{\xi}\big)-\big(m+S(r)\big)^2\right]\Psi(\vec{r},t)=0,
\end{eqnarray}
Here, $g$ is the determinant of the metric tensor, where $\sqrt{-g}=\alpha r$, $m$ is the mass and $S(r)$ is the scalar potential \cite{6,7}. The covariant derivative $D_{\zeta}$ takes the vector potential interaction into account. Note that a scalar potential is usually observed under the static field conditions whereas a vector potential is observed under the dynamic conditions. Therefore, a scalar potential is regarded as an extra quantity added to the particle mass energy.  In a minimal coupling formalism, the covariant derivative is given as
\begin{eqnarray}\label{4}
D_{\zeta}=\partial_{\zeta}- i e A_{\zeta}.
\end{eqnarray}
Here, $e$ denotes the electric charge. We take the electromagnetic four-vector potential, $A_{\zeta}$,  in the form of \cite{2,3,4,5,H4,RL,AL,BouHou}
\begin{eqnarray}\label{5}
A_{\zeta}=\left(A_{0},0 ,A_{\varphi},0\right).
\end{eqnarray}
and substitute it in Eq. \eqref{3}. We obtain
\begin{eqnarray}\label{8}
\bigg[-\Big(\partial_{t}-ieA_{0}\Big)^{2}+\frac{1}{r}\partial_{r}\Big(r\partial_{r}\Big)+\frac{1}{\alpha^{2}r^{2}}\Big(\partial_{\varphi}-ieA_{\varphi}
-\chi\partial_{z}\Big)^{2}+\partial^{2}_{z}-\Big(m+S(r)\Big)^{2}\bigg]\Psi=0
\end{eqnarray}
Note that we assume $c=\hbar=1$. Then, we follow Mirza \emph{et al.} \cite{Mirza} and introduce a  change in the momentum operator via $\hat{p}_{\zeta} \longrightarrow \hat{p}_{\zeta}+im\Omega \hat{X}_{\zeta}$. Here $\Omega$ is the oscillator frequency and $\hat{X}_{\zeta}=\big(0, f(r),0,0\big)$ \cite{2,11}. Then, we obtain the generalized K-G equation out of  Eq. \eqref{8} as
\begin{eqnarray}\label{9}
\Bigg[-\Big(\partial_{t}-ieA_{0}\Big)^{2}+\frac{1}{r}\Big(\partial_{r}+m\Omega f(r)\Big)r\Big(\partial_{r}-m\Omega f(r)\Big)+\frac{1}{\alpha^{2}r^{2}}\Big(\partial_{\varphi}-ieA_{\varphi}-\chi\partial_{z}\Big)^{2}+\partial^{2}_{z}&& \nonumber \\
-\Big(m+S(r)\Big)^{2}\Bigg]\Psi=0&&
\end{eqnarray}
Next, we consider the non-zero components of the four-vector. We choose \cite{98a}
\begin{eqnarray}\label{a0}
  eA_{0}(r) &=& V(r)
\end{eqnarray}
and
\begin{eqnarray}\label{6}
A_{\varphi}=-\frac{\alpha B_{0}}{2}r^{2}+\frac{\Phi_{B}}{2\pi }
\end{eqnarray}
Here, $\Phi_{B}$  is an internal quantum magnetic flux and it is assumed to have a constant value \cite{CF}. $B_0$ is the strength of the external magnetic field. We employ the definition of the cyclotron frequency, $\omega_{c} \equiv \frac{eB_{0}}{2 m}$, and  $\Phi\equiv \frac{e\Phi_{B}}{2\pi}$ to  rewrite $A_{\varphi}$ in the form of
\begin{eqnarray}\label{7}
-ieA_{\varphi}= i\alpha \omega_{c} m r^2-i\Phi
\end{eqnarray}
We make an ansatz and  assume that the wave function is in the form of
\begin{eqnarray}\label{10}
\Psi(t,r,\varphi,z)=e^{i\left(-Et+\ell \varphi+kz\right) } \psi(r),
\end{eqnarray}
where  $k$ is the wave number along the $\hat{e}_z$ direction, $\ell=0, \pm1, \pm2, \cdots$ is the quantum number associated to the $z$ component of the total angular momentum, and $E_{n,\ell}$ is the energy. $\psi_{n,\ell}(r)$ is the eigenfunction in the radial coordinate. Finally, we substitute Eqs. \eqref{a0}, \eqref{7} and   \eqref{10} into Eq. \eqref{9}. We obtain the following expression
\begin{eqnarray}\label{11}
\psi_{n,\ell}''(r)+\frac{1}{r}\psi_{n,\ell}'(r)+\bigg[\Big(E_{n,\ell}-V(r)\Big)^{2}-\frac{m\Omega f(r) }{r}-m\Omega f'(r)-m^{2}\Omega^{2}f^{2}(r)-k^{2}
-\Big(m+S(r)\Big)^{2}&&\nonumber\\
-\frac{1}{\alpha^{2}r^{2}}\Big(\ell - \Phi-k\chi\Big)^{2}-\frac{2m\omega_{c}}{\alpha}\Big(\ell - \Phi-k\chi\Big)-m^{2}\omega^{2}_{c}r^{2}\bigg]\psi_{n,\ell}(r)=0.&&
\end{eqnarray}
In the following sections we solve  Eq. \eqref{11} for two different cases:
\begin{itemize}
  \item {In the first case, we consider a Cornell function for the $f(r)$ and a pseudo-harmonic  function for  $V(r) = S(r)$. We find out that in this choice the differential equation obtained from Eq. \eqref{11} may be solved by the NU method.}
  \item  In the second case, we take different Cornell functions for $f(r)$, $V(r)$ and $S(r)$ by employing different coefficients. In this case, we obtain a differential equation where the NU method can not be used since a higher-order term arises.  Therefore, we employ the QES method, which would be an accurate method,  to obtain a solution.
\end{itemize}

	\baselineskip=30 pt
\section{Interaction under a pseudo harmonic-type potential energy}\label{sec3}
We consider the non-mimimal coupling function, $f(r)$, as the Cornell potential. The Cornell potential energy is defined by the linear superposition of the Coulomb-type and linear-type potential energy terms.  In short and large ranges it is dominated by Coulomb-like and linear-type terms, respectively \cite{C1, C2, C3}. It is often used to describe the binding states of heavy quarks \cite{C4, C5, C6}.
\begin{subequations}\label{12}
\begin{equation}\label{12-1}
f(r)=br +\frac{d}{r}
\end{equation}
Here, we  intend to examine the  pseudo harmonic potential energy which is a very well-know potential energy in the chemical and molecular physics \cite{C7, C8}. It is often employed in the studies that are related with the diatomic molecules \cite{C9, C10, C11}. Alike the Cornell potential, in short and long ranges it is dominated by inverse square-type and harmonic-type terms, respectively. We assume that the vector and scalar potential has the same form \cite{42,43,44,45}
\begin{equation}\label{12-2}
V(r)=S(r)=a_{1}r^{2}+\frac{a_{2}}{r^{2}}+a_{3}
\end{equation}
\end{subequations}	
where $b$, $d$, $a_{1}$, $a_{2}$ and $a_{3}$ are real constants. By substituting Eqs. \eqref{12} into Eq. \eqref{11}, we obtain
\begin{eqnarray}\label{13}
  \psi_{n,\ell}''(r)+\frac{\psi_{n,\ell}'(r)}{r}+\bigg[\xi_{1}r^2+\frac{\xi_{2}}{r^2}+\xi_{3}\bigg]\psi_{n,\ell} &=& 0
\end{eqnarray}
where $\xi_{1}, \xi_{2}$ and $\xi_{3}$ are
\begin{subequations}\label{16}
\begin{equation}
\xi_{1}=-\bigg(2a_{1}(E_{n,\ell}+m)+m^{2}\Omega^{2} b^{2}+ m^{2}\omega^{2}_{c}\bigg),
\end{equation}
\begin{equation}
\xi_{2}=-\bigg(2a_{2}(E_{n,\ell}+m)+m^{2}\Omega^{2}d^{2}+\frac{1}{\alpha^{2}}(\ell-\Phi-k\chi)^{2}\bigg),
\end{equation}
\begin{equation}
\xi_{3}=\bigg(E_{n,\ell}^{2}-k^{2}-m^{2}-2a_{3}(E_{n,\ell}+m)-2m\Omega b-2m^{2}\Omega^{2}bd-\frac{2m\omega_{c}}{\alpha}(\ell-\Phi-k\chi)\bigg).
\end{equation}
\end{subequations}
We define a new function $\psi_{n,\ell}(r)$
\begin{eqnarray}\label{14}
\psi_{n,\ell}(r) \equiv \frac{1}{\sqrt{r}}\varphi_{n,\ell}(r).
\end{eqnarray}
Then, Eq. \eqref{13} deduces to the form of
\begin{eqnarray}\label{15}
&&\varphi_{n,\ell}''(r)+\bigg[\xi_{1}r^{2}+\frac{\xi_{2}+\frac{1}{4}}{r^{2}}+\xi_{3}\bigg]\varphi_{n,\ell}(r)=0
\end{eqnarray}
Next, we consider a change of variable, $s=r^{2}$. We find
\begin{eqnarray}\label{17}
&&\varphi_{n,\ell}''(s)+\frac{1}{2s}\varphi_{n,\ell}'(s)+\frac{1}{s^{2}}\bigg[\frac{\xi_{1}}{4}s^{2}+\frac{\xi_{3}}{4}s+\frac{1}{4}
\left(\xi_{2}+\frac{1}{4}\right)\bigg]\varphi_{n,\ell}(s)=0
\end{eqnarray}
We employ the NU method \cite{MogtignyZare2018}, where the second order differential equation has the form of
\begin{eqnarray} \label{NUform}
  \phi''(s)+\bigg[\frac{\alpha_1-\alpha_2 s}{s(1-\alpha_3 s)}\bigg]\phi'(s)+\bigg[\frac{-\beta_1s^2+\beta_2s-\beta_3}{s^2(1-\alpha_3 s)}\bigg] \phi(s)&=& 0
\end{eqnarray}
and compare Eq. \eqref{17} with Eq. \eqref{NUform}. We find the coefficients as
\begin{eqnarray}\label{18}
&&\alpha_{1}=\frac{1}{2},\,\,\,\,\, \alpha_{2}=\alpha_{3}=0,\,\,\,\,\,\beta_{1}=-\frac{\xi_{1}}{4},\,\,\,\,\,\beta_{2}=\frac{\xi_{3}}{4},\,\,\,\,\, \beta_{3}=- \frac{1}{4}\left(\xi_{2}+\frac{1}{4}\right), 
\end{eqnarray}
Then, we substitute the coefficients to the  algorithm of the NU method and we obtain the wave function as
\begin{eqnarray}\label{18+1}
\psi_{n,\ell}(r)=N_{n,\ell}r^{\sqrt{-\xi_2}}e^{-\frac{\sqrt{-\xi_{1}}}{2} r^{2}}L_{n}^{\sqrt{-\xi_2}}(\sqrt{-\xi_{1}}r^{2}).
\end{eqnarray}
Here, $N_{n,\ell}$ is the normalization constant. $L_{n}^{\alpha}(x)$ represents the generalized Laguerre polynomial while $n$ denotes the quantum number. We derive an analytic expression to calculate the energy egienvalues as follows
\begin{eqnarray}\label{183}
\xi_{3}&=&2\sqrt{-\xi_{1}}\Big[\sqrt{-\xi_{2}}+(2n+1)\Big].
\end{eqnarray}
We substitute Eq. \eqref{16} in Eq. \eqref{183}. We find
\begin{eqnarray}\label{19}
E_{n,\ell}^{2}&=&k^{2}+m^{2}+2a_{3}(E_{n,\ell}+m)+2m\Omega b+2m^{2}\Omega^{2}bd+\frac{2m\omega_{c}}{\alpha}(\ell-\Phi-k\chi)\\
&+&2\sqrt{2a_{1}(E_{n,\ell}+m)+m^{2}\Omega^{2} b^{2}+ m^{2}\omega^{2}_{c}} \bigg[(2n+1) +\sqrt{2a_{2}(E_{n,\ell}+m)+m^{2}\Omega^{2}d^{2}+\frac{1}{\alpha^{2}}(\ell-\Phi-k\chi)^{2}}\bigg].\nonumber
\end{eqnarray}
where $n=0,1,2,\cdots$.

\begin{figure}[ht!]
	\centering
	\includegraphics[width=0.8\textwidth]{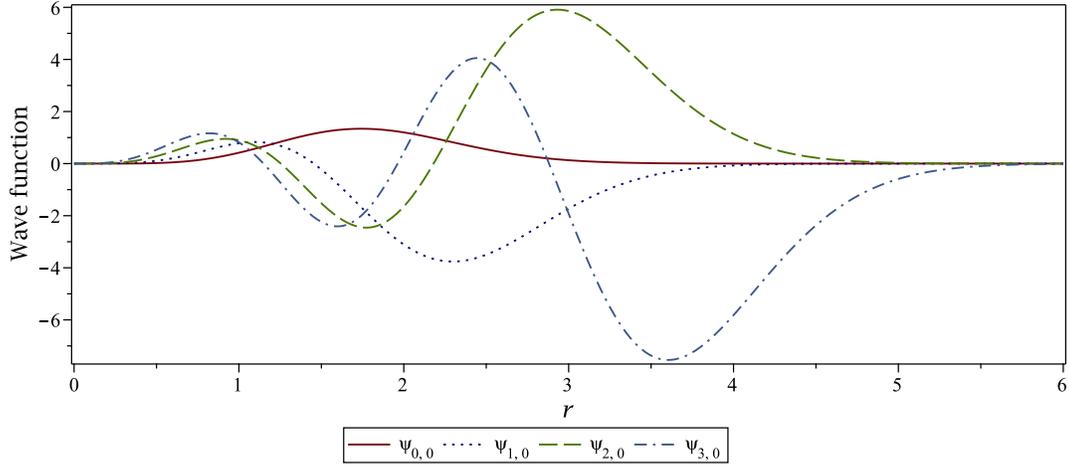}
	\caption{Unnormalized wave functions versus the distance.}
	\label{fig:wf1}
\end{figure}

We take $a_1=a_2=a_3=b=d=k=\chi=\omega_c=\Phi=1$, $\alpha=0.8$, $\Omega=0.2$, $m=2$ and evaluate some of the energy eigenvalues to demonstrate the corresponding wave functions. We present our result in Fig. \ref{fig:wf1}.

\begin{figure}
\centering
\begin{subfigure}{.5\textwidth}
  \centering
  \includegraphics[width=\linewidth]{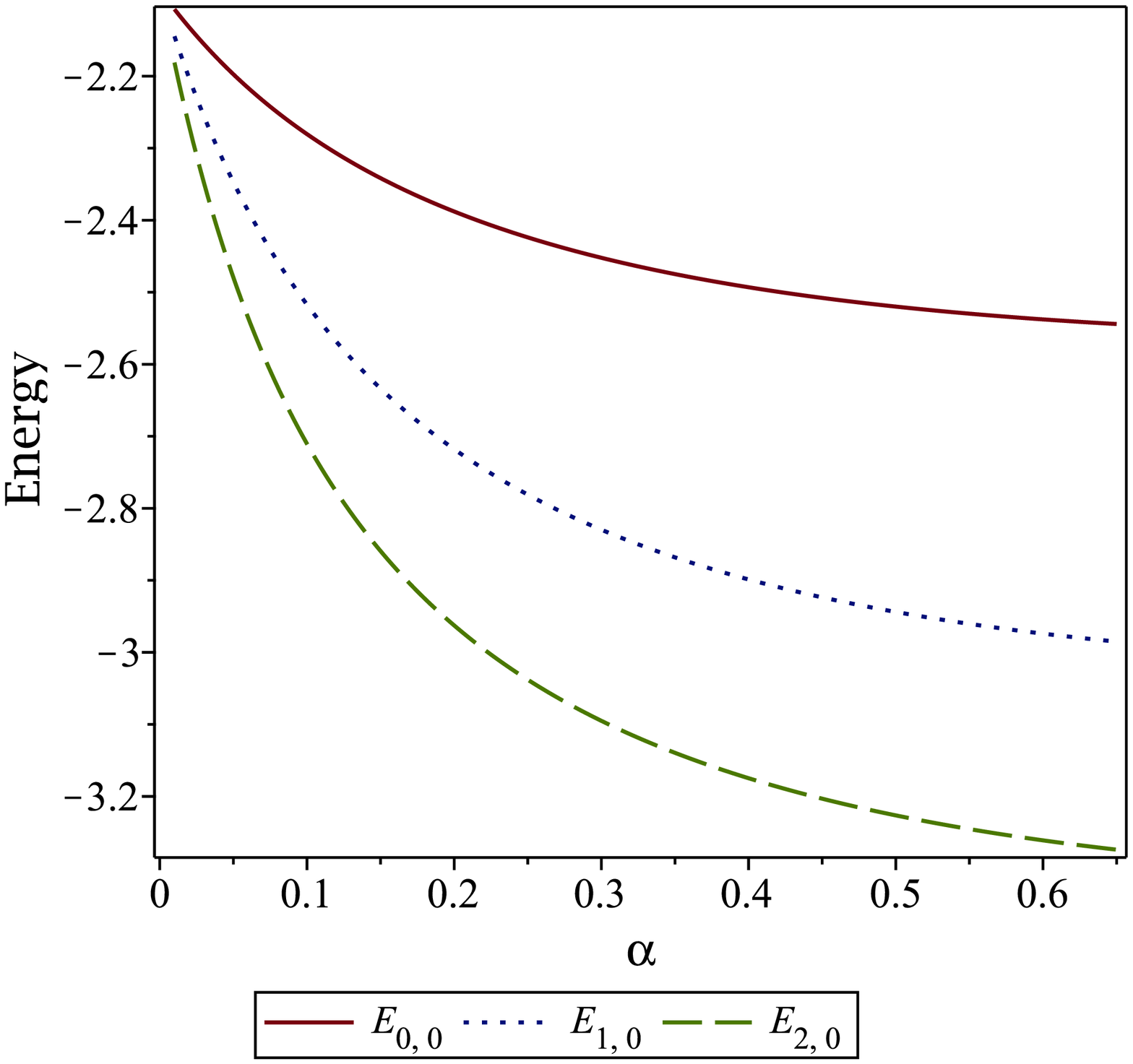}
  \caption{Ground state with the first excited two states \\ for zero angular momentum.}
  \label{fig:BCLPHa1}
\end{subfigure}%
\begin{subfigure}{.5\textwidth}
  \centering
  \includegraphics[width=\linewidth]{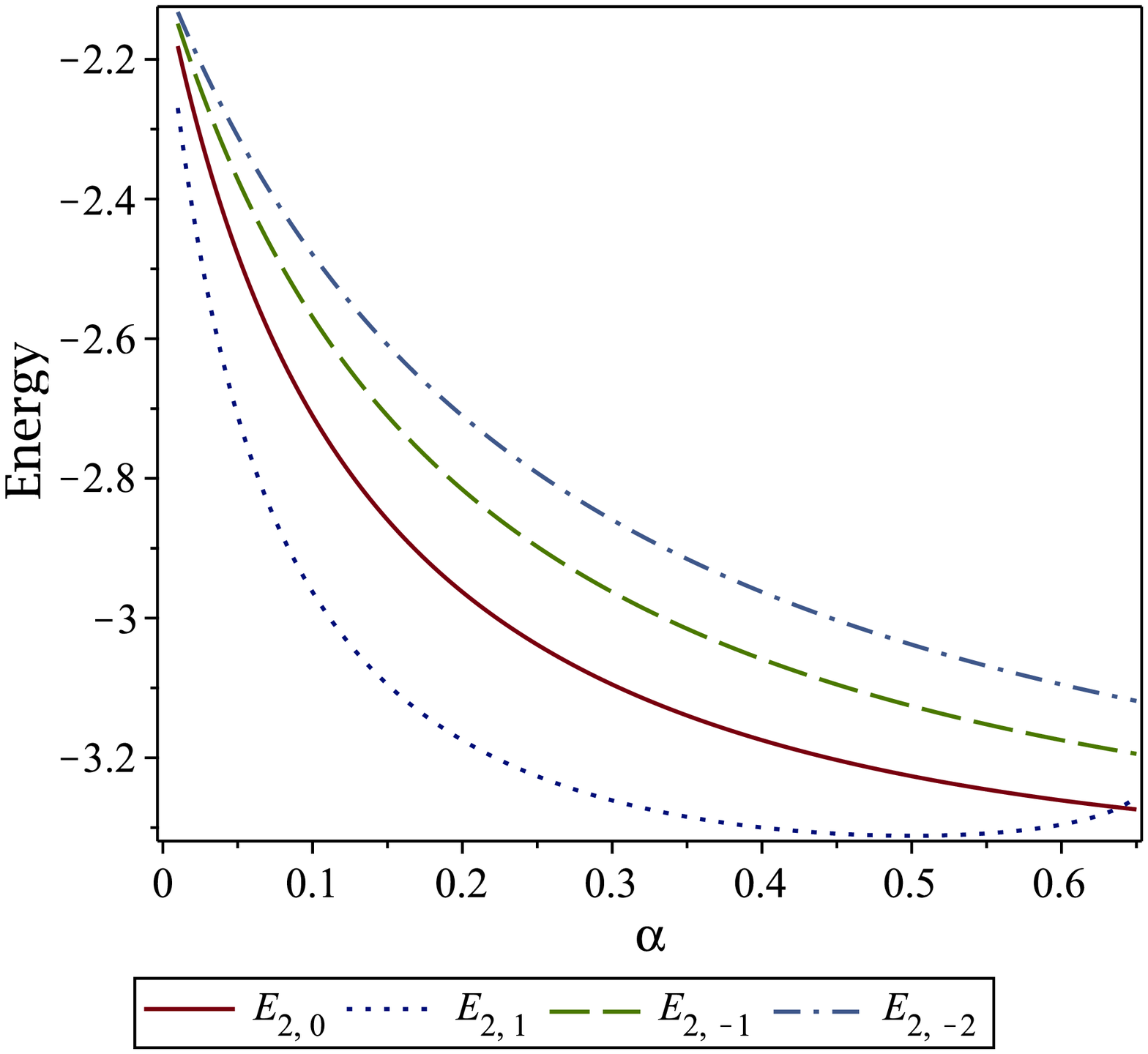}
  \caption{Second excited state with the non-zero angular momentum quantum numbers.}
  \label{fig:BCLPHa2}
\end{subfigure}
\caption{The energy spectrum function versus the angular parameter for $\Omega=0.2$ and $\Phi=\chi=1$.}
\label{fig:BCLPHalpha}
\end{figure}

\begin{figure}
\centering
\begin{subfigure}{.5\textwidth}
  \centering
  \includegraphics[width=\linewidth]{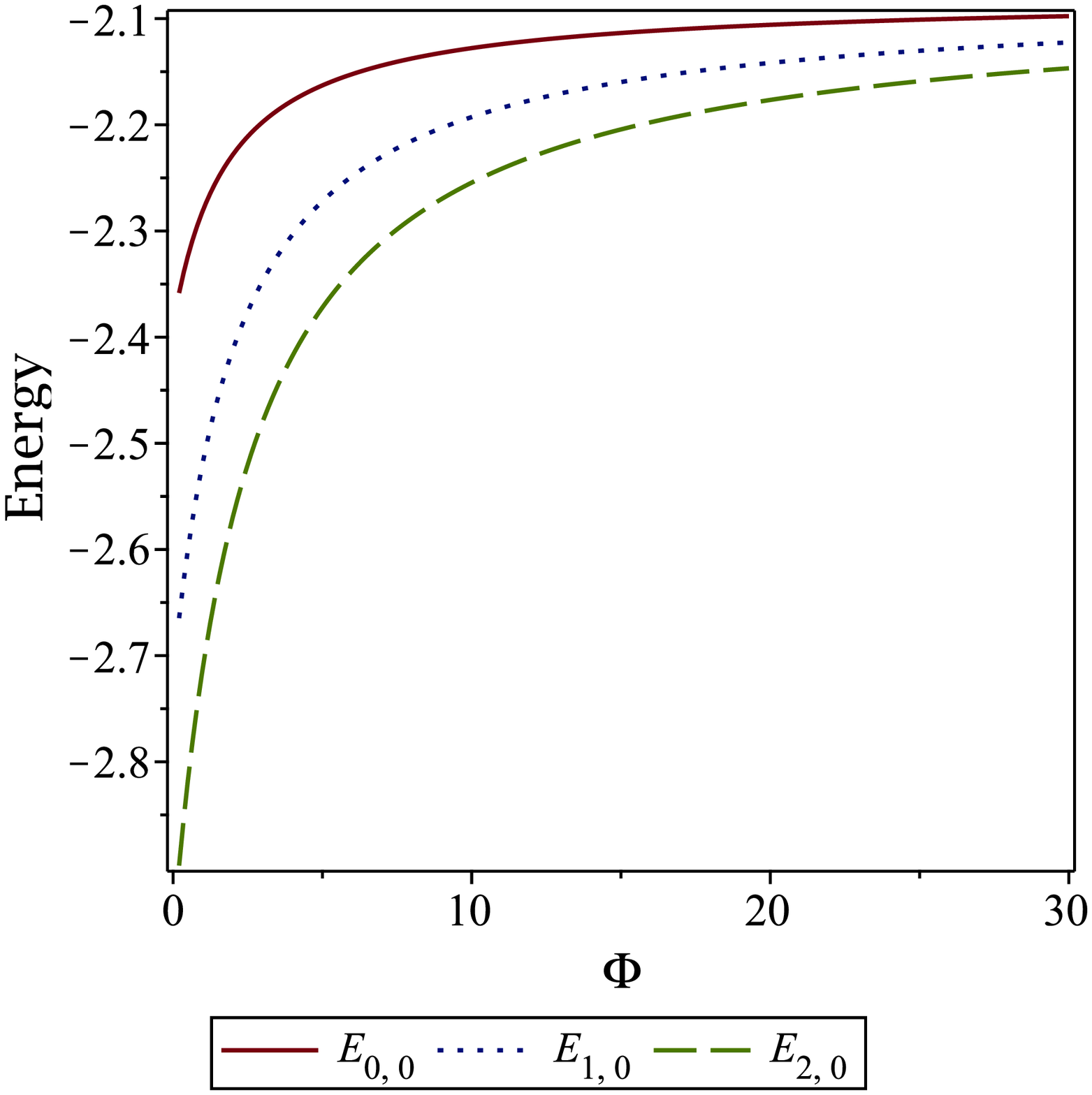}
  \caption{Ground state with the first excited two states \\ for zero angular momentum.}
  \label{fig:BCLPHp1}
\end{subfigure}%
\begin{subfigure}{.5\textwidth}
  \centering
  \includegraphics[width=\linewidth]{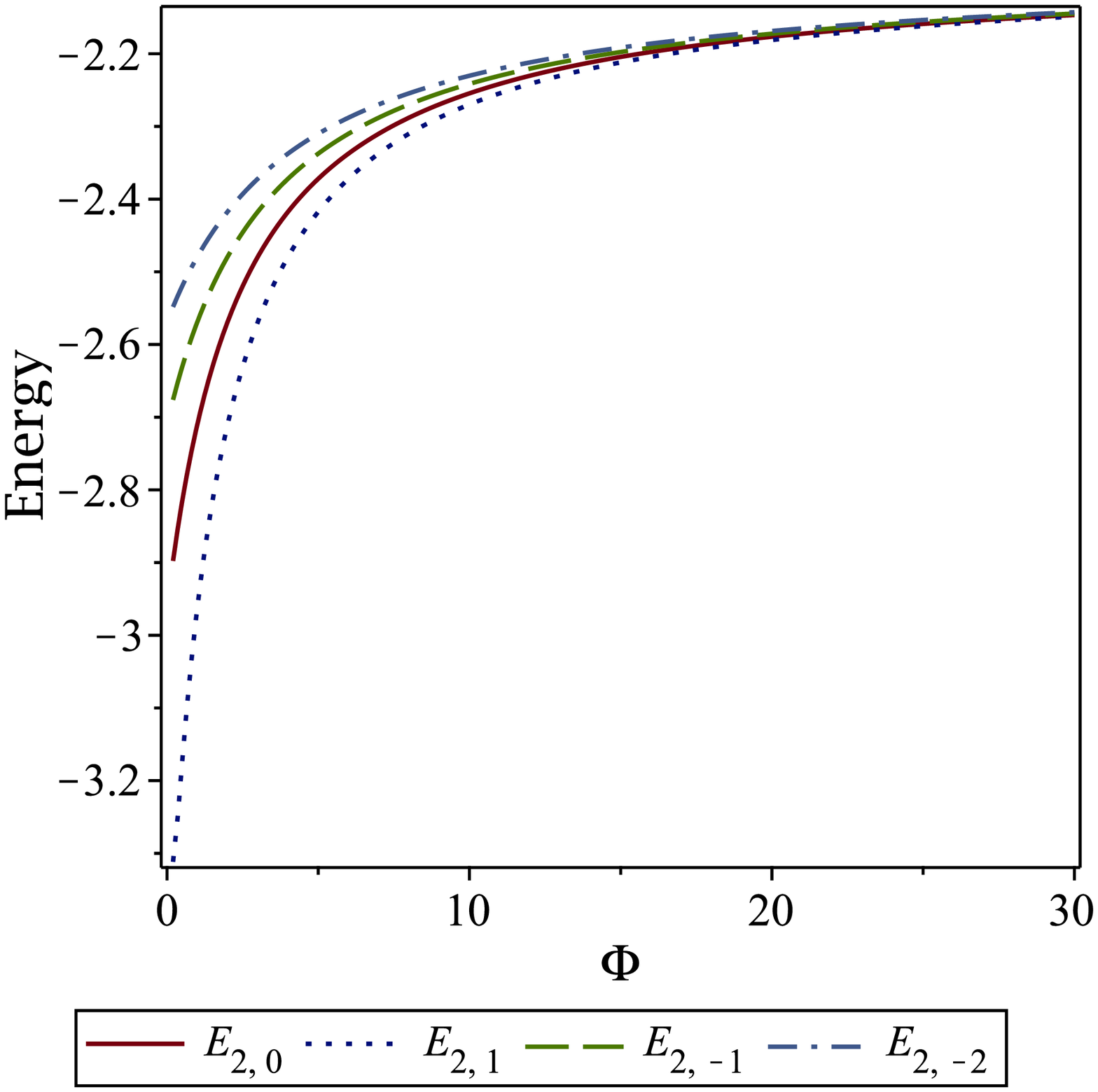}
  \caption{Second excited state with the non-zero angular momentum quantum numbers.}
  \label{fig:BCLPHp2}
\end{subfigure}
\caption{The energy spectrum function versus the angular parameter for $\alpha=0.1$, $\Omega=0.2$, and $\chi=1$.}
\label{fig:BCLPHphi}
\end{figure}

\begin{figure}
\centering
\begin{subfigure}{.5\textwidth}
  \centering
  \includegraphics[width=\linewidth]{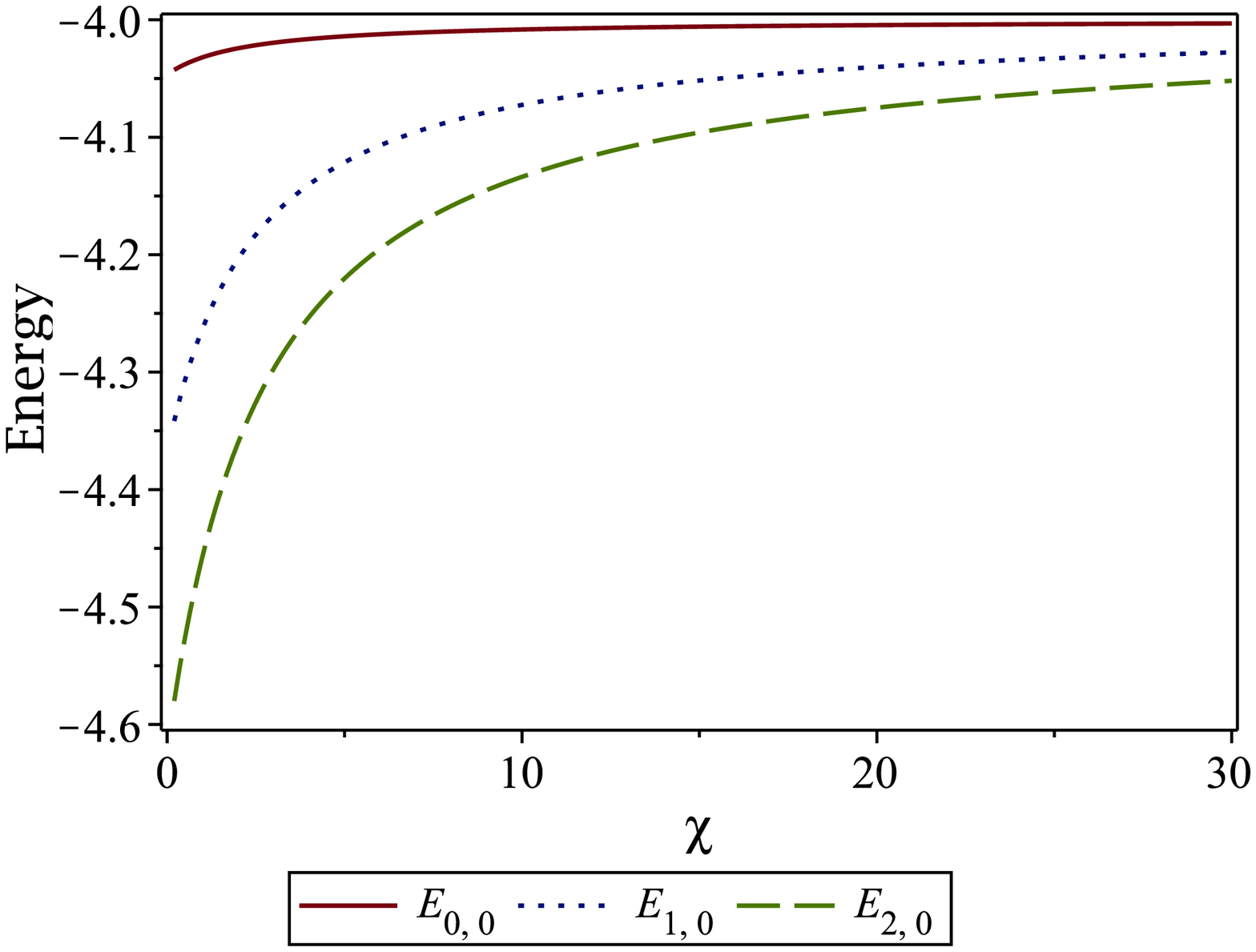}
  \caption{Ground state with the first excited two states \\ for zero angular momentum.}
  \label{fig:BCLPHc1}
\end{subfigure}%
\begin{subfigure}{.5\textwidth}
  \centering
  \includegraphics[width=\linewidth]{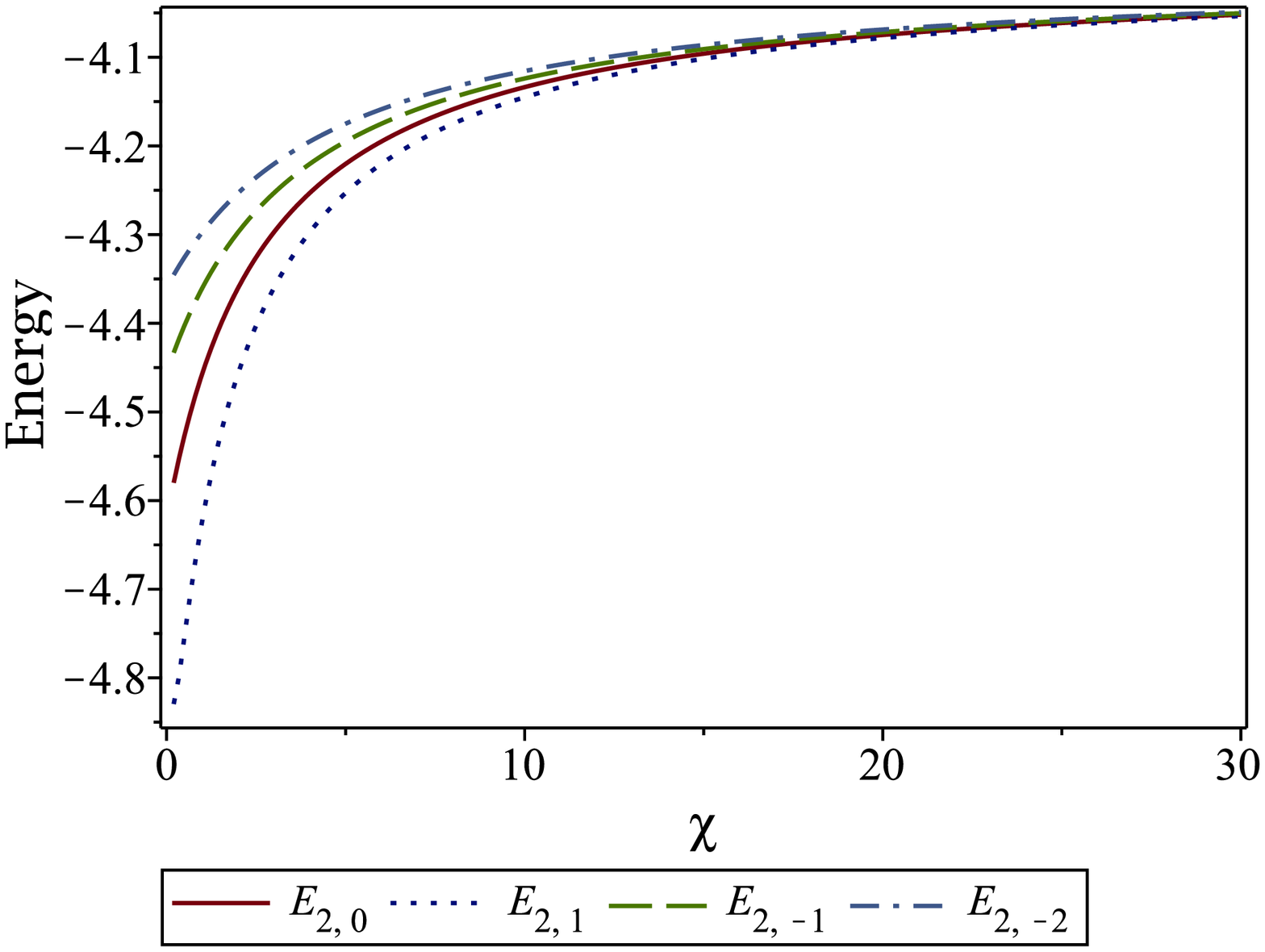}
  \caption{Second excited state with the non-zero angular momentum quantum numbers.}
  \label{fig:BCLPHc2}
\end{subfigure}
\caption{The energy spectrum function versus the angular parameter for $\alpha=0.1$, $\Omega=1$, and $\Phi=1$.}
\label{fig:BCLPHchi}
\end{figure}

We explore the effect of the angular parameter, magnetic flux and the torsion parameters on the energy spectrum. We take $a_1=a_2=a_3=b=d=k=\omega=1$,  and $m=2$. We  plot the energy eigenfunction versus $\alpha$, $\Phi$, $\chi$ parameters in Figs. \ref{fig:BCLPHalpha},  \ref{fig:BCLPHphi}, and \ref{fig:BCLPHchi}, respectively.

\subsection{Linearly non-minimal coupling limit}
In this limit, we assume that the pseudo harmonic interaction does not exist. Therefore, we take $a_1=a_2=a_3=0$. Moreover, we consider a linear non-minimal coupling, hence, we use $d=0$ and $b=1$. We find
\begin{eqnarray}\label{19LCL}
E_{n,\ell}^{2}&=&k^{2}+m^{2}+2m\Omega +\frac{2m\omega_{c}}{\alpha}(\ell-\Phi-k\chi)+2m\sqrt{\Omega^{2}+ \omega^{2}_{c}} \bigg[(2n+1) + \frac{|\ell-\Phi-k\chi|}{\alpha}\bigg].
\end{eqnarray}
In the zero torsion limit, Eq. \eqref{19LCL} yields to the same result that is given in Eq. $(27)$ of \cite{2}.

\subsection{Minimal coupling limit}
When $\Omega$ tends to zero, we obtain the minimal coupling limit. Then, the energy spectrum function that is derived in Eq. \eqref{19} reduces to
\begin{eqnarray}\label{19MCL}
E_{n,\ell}^{2}&=&k^{2}+m^{2}+2a_{3}(E_{n,\ell}+m)+\frac{2m\omega_{c}}{\alpha}(\ell-\Phi-k\chi)\nonumber \\
&+&2\sqrt{2a_{1}(E_{n,\ell}+m)+ m^{2}\omega^{2}_{c}} \bigg[(2n+1) +\sqrt{2a_{2}(E_{n,\ell}+m)+\frac{1}{\alpha^{2}}(\ell-\Phi-k\chi)^{2}}\bigg].
\end{eqnarray}
In this limit, if we consider the external magnetic field does not exist, we obtain
\begin{eqnarray}\label{19MCL2}
E_{n,\ell}^{2}=k^{2}+m^{2}+2a_{3}(E_{n,\ell}+m)+2\sqrt{2a_{1}(E_{n,\ell}+m)} \bigg[(2n+1) +\sqrt{2a_{2}(E_{n,\ell}+m)+\frac{1}{\alpha^{2}}(\ell-\Phi-k\chi)^{2}}\bigg].
\end{eqnarray}
\subsection{Zero magnetic field limit}
When we examine the zero external magnetic field in the general form of Eq. \eqref{19}, we find
\begin{eqnarray}\label{19ZMF}
E_{n,\ell}^{2}&=&k^{2}+m^{2}+2a_{3}(E_{n,\ell}+m)+2m\Omega b+2m^{2}\Omega^{2}bd\nonumber \\
&+&2\sqrt{2a_{1}(E_{n,\ell}+m)+m^{2}\Omega^{2} b^{2}} \bigg[(2n+1) +\sqrt{2a_{2}(E_{n,\ell}+m)+m^{2}\Omega^{2}d^{2}+\frac{1}{\alpha^{2}}(\ell-\Phi-k\chi)^{2}}\bigg].
\end{eqnarray}
If we consider the zero internal magnetic flux case, we get
\begin{eqnarray}\label{19ZMF2}
E_{n,\ell}^{2}&=&k^{2}+m^{2}+2a_{3}(E_{n,\ell}+m)+2m\Omega b+2m^{2}\Omega^{2}bd\nonumber \\
&+&2\sqrt{2a_{1}(E_{n,\ell}+m)+m^{2}\Omega^{2} b^{2}} \bigg[(2n+1) +\sqrt{2a_{2}(E_{n,\ell}+m)+m^{2}\Omega^{2}d^{2}+\frac{1}{\alpha^{2}}(\ell-k\chi)^{2}}\bigg].
\end{eqnarray}

\subsection{Harmonic-type potential energy limit}
In $a_{2}, a_{3}\longrightarrow 0$  limit, the vector, thus scalar potential energies reduce to the harmonic oscillator potential energy. Therefore, in this limit Eq. \eqref{19} reduces to
\begin{eqnarray}\label{20}
E_{n,\ell}^{2}&=&k^{2}+m^{2}+2m\Omega b+2m^{2}\Omega^{2}bd+\frac{2m\omega_{c}}{\alpha}(\ell-\Phi-k\chi)\nonumber\\
&+&2\sqrt{2a_{1}(E_{n,\ell}+m)+m^{2}\Omega^{2} b^{2}+ m^{2}\omega^{2}_{c}} \bigg[(2n+1) +\sqrt{m^{2}\Omega^{2}d^{2}+\frac{1}{\alpha^{2}}(\ell-\Phi-k\chi)^{2}}\bigg].
\end{eqnarray}
Next we consider the minimal coupling limit, namely $\Omega \rightarrow 0$, then the energy function becomes
\begin{eqnarray}\label{20a}
E_{n,\ell}^{2}&=&k^{2}+m^{2}+\frac{2m\omega_{c}}{\alpha}(\ell-\Phi-k\chi)+2\sqrt{2a_{1}(E_{n,\ell}+m)+ m^{2}\omega^{2}_{c}} \bigg[(2n+1) +\frac{|\ell-\Phi-k\chi|}{\alpha}\bigg].
\end{eqnarray}
If we assume zero internal magnetic flux and zero spatial dislocation, then  Eq. \eqref{20a} reduces to
\begin{eqnarray}\label{21}
E_{n,\ell}^{2}&=&k^{2}+m^{2}+\frac{2m\omega_{c}\ell}{\alpha}+2\sqrt{2a_{1}(E_{n,\ell}+m)+ m^{2}\omega^{2}_{c}} \bigg[(2n+1) +\frac{|\ell|}{\alpha}\bigg].
\end{eqnarray}
\subsection{Inverse square-type potential energy limit}
In this limit, we consider $a_{1}, a_{3}\longrightarrow 0$, that is, there is only inverse square term exists in the vector and scalar potential.
Then,  Eq. \eqref{19} becomes
\begin{eqnarray}\label{22}
E_{n,\ell}^{2}&=&k^{2}+m^{2}+2m\Omega b+2m^{2}\Omega^{2}bd+\frac{2m\omega_{c}}{\alpha}(\ell-\Phi-k\chi)\\
&+&2\sqrt{m^{2}\Omega^{2} b^{2}+ m^{2}\omega^{2}_{c}} \bigg[(2n+1) +\sqrt{2a_{2}(E_{n,\ell}+m)+m^{2}\Omega^{2}d^{2}+\frac{1}{\alpha^{2}}(\ell-\Phi-k\chi)^{2}}\bigg].\nonumber
\end{eqnarray}
In the minimal coupling limit, Eq. \eqref{22} reduces to
\begin{eqnarray}\label{22a}
E_{n,\ell}^{2}&=&k^{2}+m^{2}+\frac{2m\omega_{c}}{\alpha}(\ell-\Phi-k\chi)+2m\omega_{c} \bigg[(2n+1) +\sqrt{2a_{2}(E_{n,\ell}+m)+\frac{1}{\alpha^{2}}(\ell-\Phi-k\chi)^{2}}\bigg].
\end{eqnarray}
We remark that, in the case of the vanishing magnetic field, Eq. \eqref{22a} reduces to the well-known relativistic energy form. In the limit, where internal magnetic flux, and spatial dislocation terms go to zero,  Eq. \eqref{22a} becomes
\begin{eqnarray}\label{23}
E_{n,\ell}^{2}&=&k^{2}+m^{2}+\frac{2m\omega_{c}\ell}{\alpha}+2m\omega_{c} \bigg[(2n+1) +\sqrt{2a_{2}(E_{n,\ell}+m)+\frac{\ell^2}{\alpha^{2}}}\bigg].
\end{eqnarray}
\subsection{AB effect}
In all cases, where $\Phi_{B}\neq0$ and $\chi\neq 0$, a change in the magnetic quantum number can be compensated with a change of the magnetic flux.
\begin{eqnarray}\label{28+22}
E_{n,\ell}\Big(\Phi_{B}\pm  \frac{2\pi}{e}\eta\Big)=E_{n,\ell\mp \eta }\Big(\Phi_{B}\Big),
\end{eqnarray}
where $\eta=1,2,3,\cdots$. We observe that the torsion of space-time plays a role in the determination of the values of the energy spectrum \cite{51a}.  Although the AB effect is characterized by the topological constant, magnetic flux and torsion, we realize that the AB effect is valid even in the case of zero torsion \cite{2}.
\subsection{Degeneracy}
We observe degenerate states. For example, in the minimal coupling and zero external magnetic filed limit, with $a_2=a_3=0$ and $\Phi=\chi=0$ values we find some of the following degenerate states out of Eq. \eqref{19}.
\begin{itemize}
  \item For $\alpha=1$,
        \begin{subequations}
        \begin{eqnarray}
         E_{0,\mp 7} =  E_{1,\mp 5} =  E_{2,\mp 3} &=& E_{3,\mp 1}, \\
                        E_{0,\mp 5} =  E_{1,\mp 3} &=& E_{2,\mp 1}, \\
                                       E_{0,\mp 3} &=& E_{1,\mp 1}.
         \end{eqnarray}
         \end{subequations}
  \item For $\alpha=\frac{1}{2}$,
   \begin{subequations}
   \begin{eqnarray}
    E_{0,\mp 4} =  E_{1,\mp 3} =  E_{2,\mp 2} &=& E_{3,\mp 1}, \\
                   E_{0,\mp 3} =  E_{1,\mp 2} &=& E_{2,\mp 1}, \\
                                  E_{0,\mp 2} &=& E_{1,\mp 1}.
  \end{eqnarray}
  \end{subequations}
\end{itemize}

As we know, usually an external effect removes the degeneracy or at least reduces the amount of the degeneracy states numbers. We have determined some of the degeneracy states. We have shown that when $\omega_{c}$ increases, differences between the energy eigenvalues of the degenerate states occur, and thus, the degeneracy vanishes. Also, we observe a periodic behavior for $\Phi$, as one expects since $\Phi$ is quantized. In Eq. \eqref{19}, if    $(a_{1},   a_{2}, a_{3}\longrightarrow 0)$  and  $ \Omega \longrightarrow 0$  we have
\begin{eqnarray}\label{19n}
E_{n,\ell}=\pm \sqrt{k^{2}+m^{2}+\frac{2m\omega_{c}}{\alpha}(\ell-\Phi-k\chi)+2|m\omega_{c}|\bigg(2n+1 +\bigg|\frac{\ell-\Phi-k\chi}{\alpha}\bigg|\bigg)},
\end{eqnarray}
\begin{itemize}
  \item For the case $(\ell-\Phi-k\chi)<0$, we get that $E_{n,\ell}=E_{n',\ell'}$ that means the energy eigenvalues are independent of $\ell$.
  \item For the other case, where $(\ell-\Phi-k\chi)>0$, we have degenerate states when  $n+\ell=n'+\ell'$. The increases or decreases of the parameter $\omega_{c}$ do not have any effect on degeneracy. Unlike, the increases or decreases of $\Phi$ or $\chi$ remove the degeneracy. If we consider $a_1$ and $a_2$ have non-zero values, we get dominant effect from the parameters $\Phi$, $\chi$ and $\omega_{c}$. In this case, an increase in these parameters yields a decrease in the number of degenerate states. Moreover, it is worth noting that in that case, a formal relation between the degenerate states does not exist.
\end{itemize}
\color{black}
\section{Interaction under a Cornell-type  potential energy}\label{sec4}
\begin{figure}
\centering
\begin{subfigure}{.5\textwidth}
  \centering
  \includegraphics[width=\linewidth]{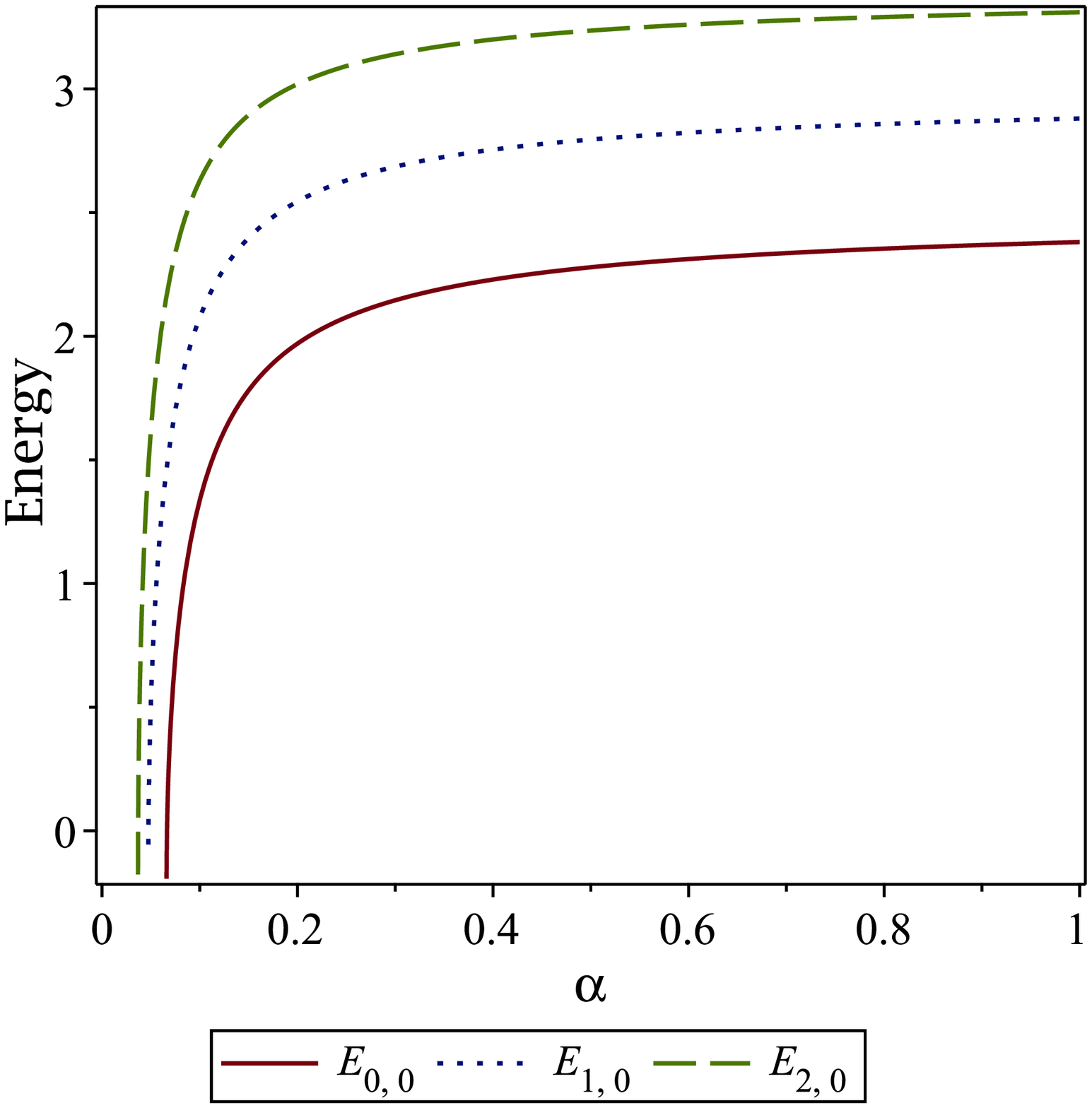}
  \caption{Ground state with the first excited two states \\ for zero angular momentum.}
  \label{fig:BCLCorna1}
\end{subfigure}%
\begin{subfigure}{.5\textwidth}
  \centering
  \includegraphics[width=\linewidth]{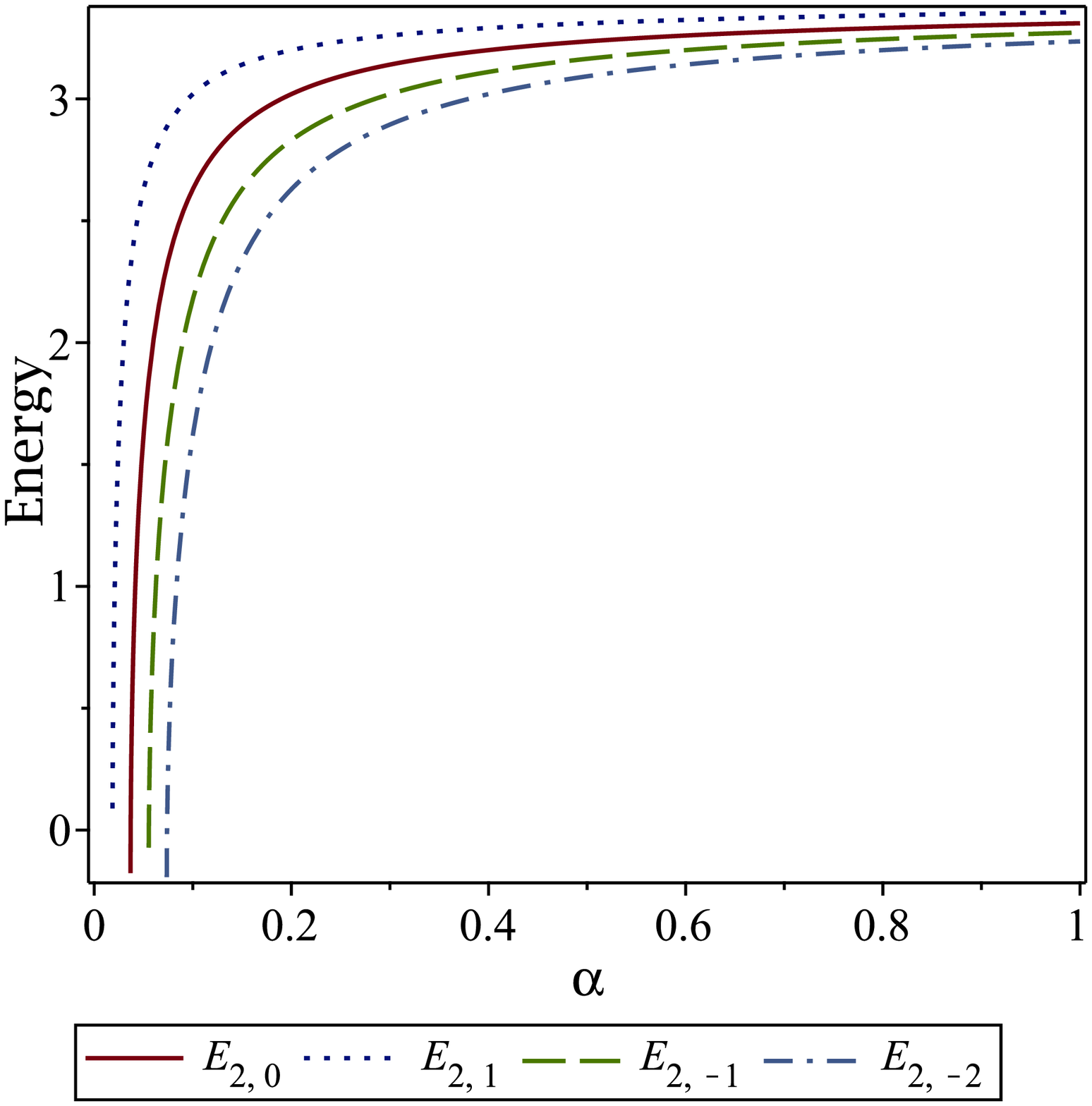}
  \caption{Second excited state with the non-zero angular momentum quantum numbers.}
  \label{fig:BCLCorna2}
\end{subfigure}
\caption{The energy spectrum function versus the angular parameter for $\Omega=0.2$ and $\Phi=\chi=1$.}
\label{fig:BCLCornalpha}
\end{figure}

\begin{figure}
\centering
\begin{subfigure}{.5\textwidth}
  \centering
  \includegraphics[width=\linewidth]{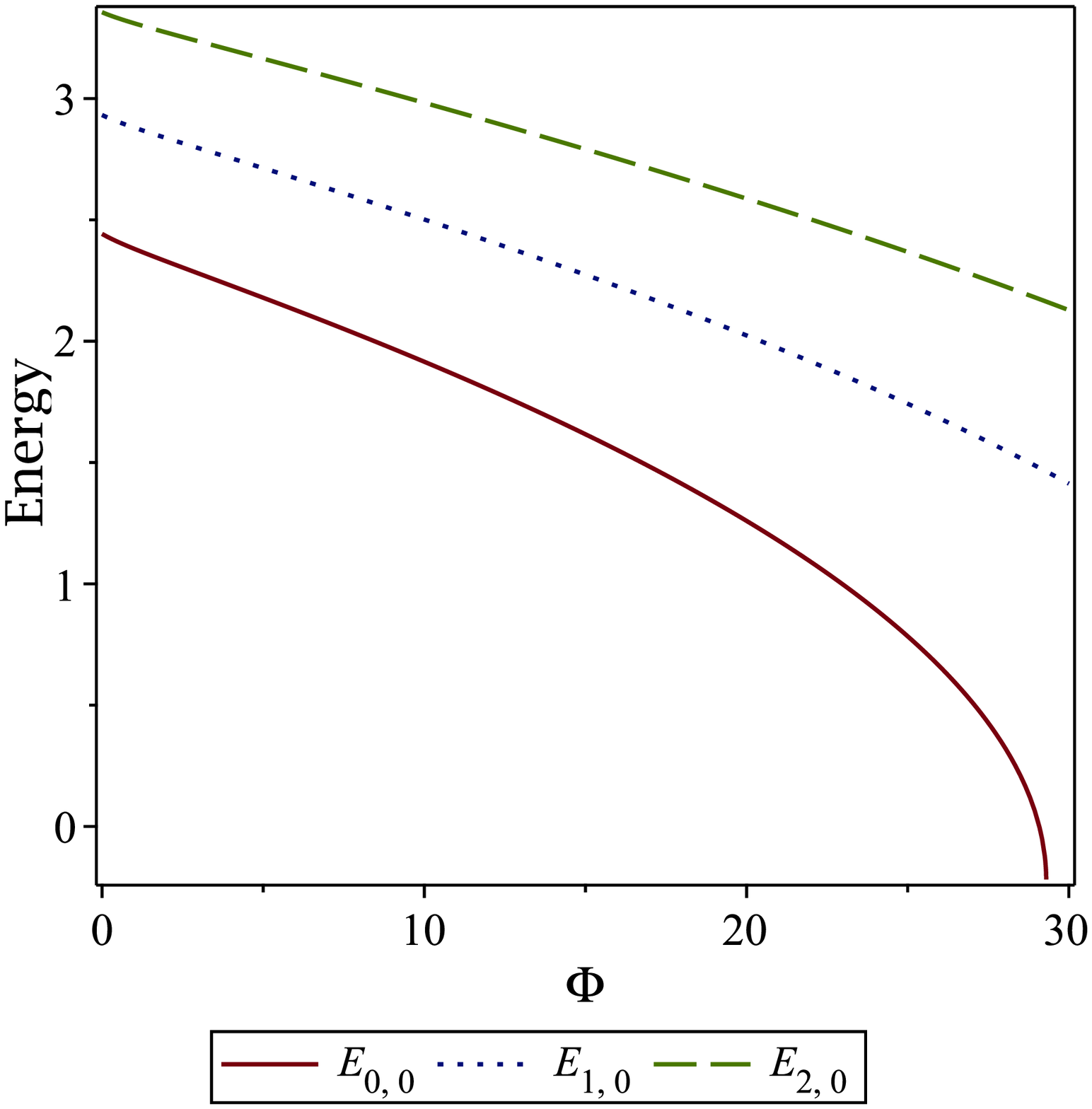}
  \caption{Ground state with the first excited two states \\ for zero angular momentum.}
  \label{fig:BCLCornp1}
\end{subfigure}%
\begin{subfigure}{.5\textwidth}
  \centering
  \includegraphics[width=\linewidth]{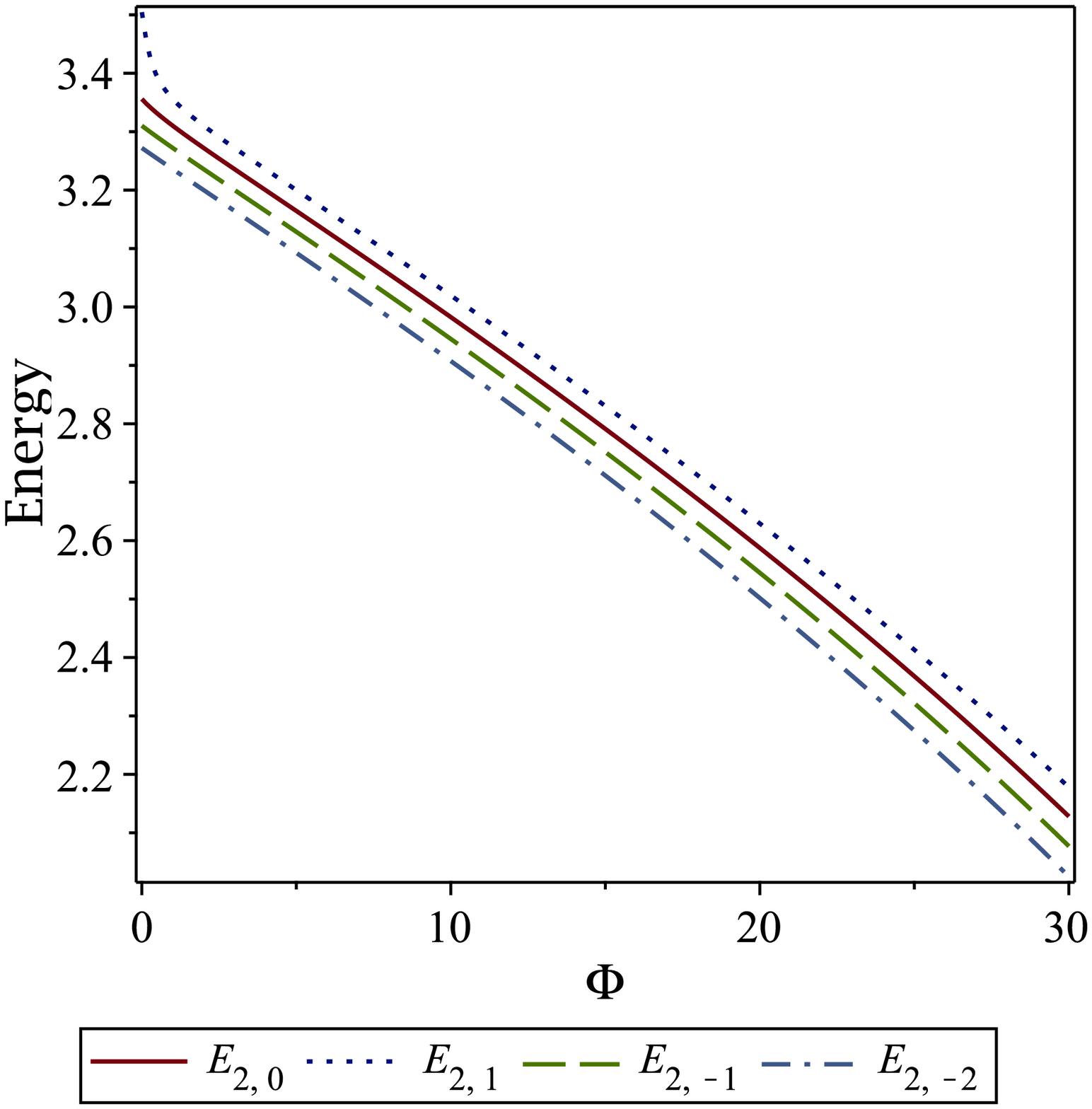}
  \caption{Second excited state with the non-zero angular momentum quantum numbers.}
  \label{fig:BCLCornp2}
\end{subfigure}
\caption{The energy spectrum function versus the angular parameter for $\alpha=1$, $\Omega=0.2$, and $\chi=1$.}
\label{fig:BCLCornphi}
\end{figure}

\begin{figure}
\centering
\begin{subfigure}{.5\textwidth}
  \centering
  \includegraphics[width=\linewidth]{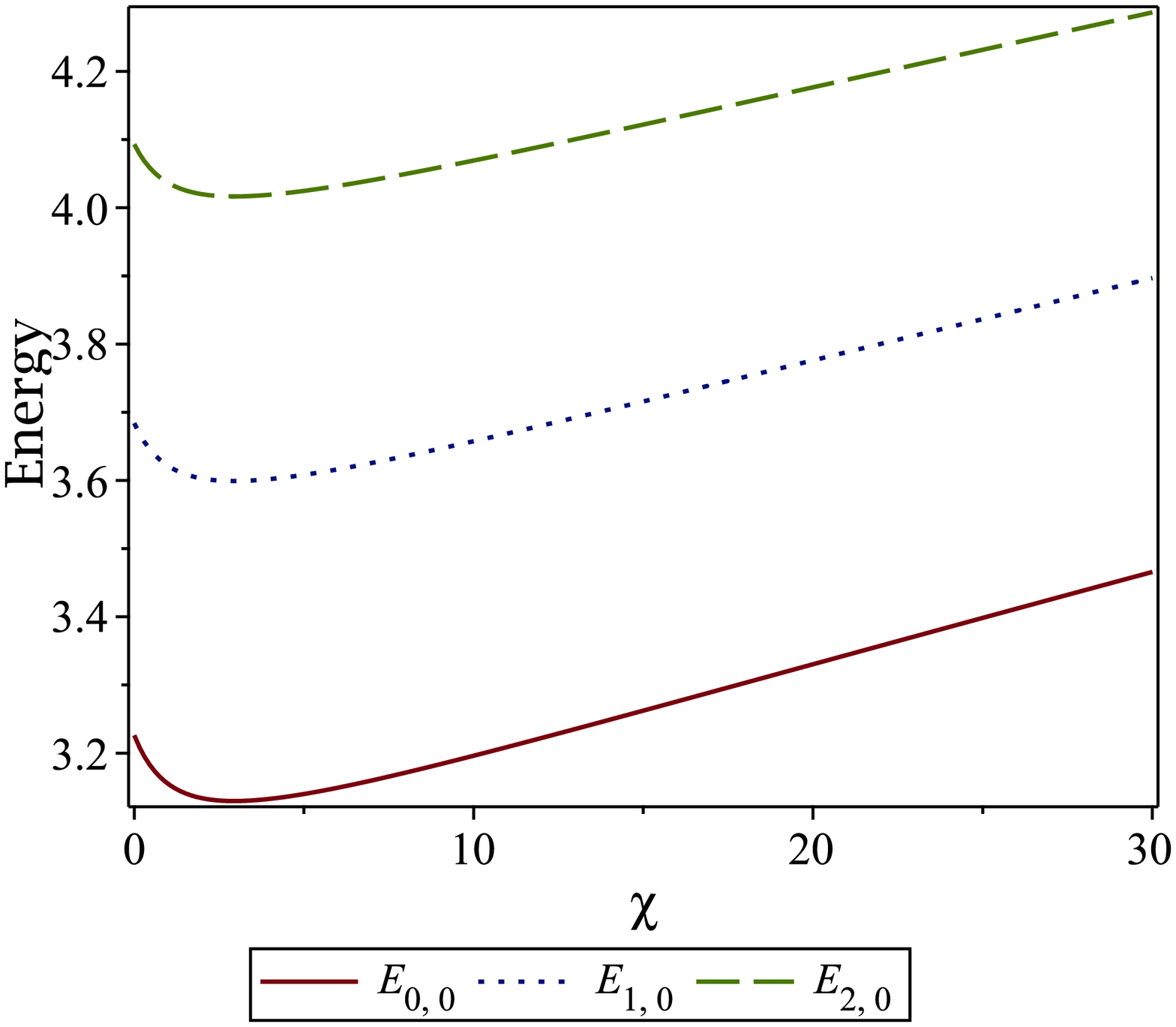}
  \caption{Ground state with the first excited two states \\ for zero angular momentum.}
  \label{fig:BCLCornc1}
\end{subfigure}%
\begin{subfigure}{.5\textwidth}
  \centering
  \includegraphics[width=\linewidth]{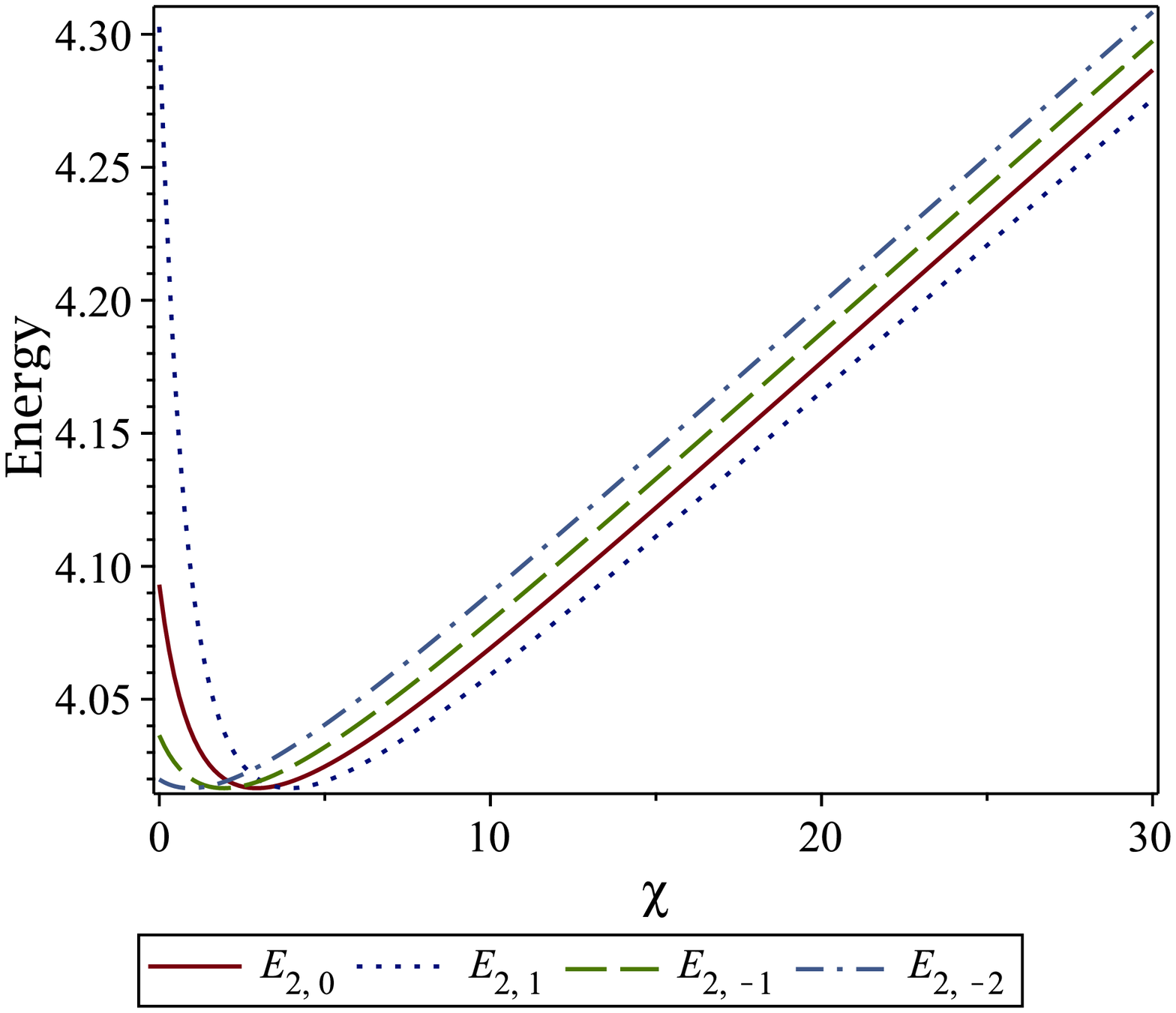}
  \caption{Second excited state with the non-zero angular momentum quantum numbers.}
  \label{fig:BCLCornc2}
\end{subfigure}
\caption{The energy spectrum function versus the angular parameter for $\alpha=1$, $\Omega=0.5$, and $\Phi=1$.}
\label{fig:BCLCornchi}
\end{figure}

In this section, we use a radial function with the Cornell potential form as performed in the previous section. However, we choose the scalar and vector potential energies from the unequal amplitude Cornell-type potential energies \cite{2,9,10}
\begin{subequations}\label{24}
\begin{eqnarray}
f(r)&=&br +\frac{d}{r} \\
V(r)&=&v_{0} r+\frac{v_{1}}{r} \\
S(r)&=&s_{0} r+\frac{s_{1}}{r}
\end{eqnarray}
\end{subequations}
where $b$, $d$, $v_{0}$, $v_{1}$, $s_{0}$ and $s_{1}$ are real constants. We substitute Eqs. \eqref{24} into Eq. \eqref{11}. We obtain a second order differential equation in the form of
\begin{eqnarray} \label{QES1}
H\psi_{n,\ell}=0
\end{eqnarray}
Here, $H$, is the Hamilton operator  given as
\begin{eqnarray}\label{25}
H=\frac{d^{2}}{dr^{2}}+\frac{1}{r}\frac{d}{dr}+\frac{\lambda_{1}}{r^{2}}+\frac{\lambda_{2}}{r}+\lambda_{3}+\lambda_{4}r+\lambda_{5}r^{2}
\end{eqnarray}
and $\lambda_{1}$, $\lambda_{2}$, $\lambda_{3}$, $\lambda_{4}$, $\lambda_{5}$  are
\begin{subequations}\label{25+1}
\begin{eqnarray}
\lambda_{1}&=&-\bigg(m^{2}\Omega^{2}d^{2}+s^{2}_{1}-v^{2}_{1}+\frac{1}{\alpha^{2}}(\ell-\Phi-k\chi)^{2}\bigg), \\
\lambda_{2}&=&-2\bigg(E_{n,\ell}v_{1}+ms_{1}\bigg),\\
\lambda_{3}&=&\bigg(E_{n,\ell}^{2}-k^{2}-m^{2}-2m\Omega b-2m^{2}\Omega^{2}bd- 2(s_{0}s_{1}-v_{0}v_{1})- \frac{2m\omega_{c}}{\alpha} (\ell-\Phi-k\chi) \bigg),\\
\lambda_{4}&=&-2\bigg(E_{n,\ell}v_{0}+ms_{0}\bigg),\\
\lambda_{5}&=&-\bigg(m^{2}\Omega^{2} b^{2}+ m^{2}\omega^{2}_{c}+s^{2}_{0}-v^{2}_{0}\bigg).
\end{eqnarray}
\end{subequations}
We use the QES technique  which is based on a gauge transformation that transforms the Hamilton operator and the wave function \cite{Seda}. To be more specific, we employ a gauge operator $G$ for transforming Eq. \eqref{QES1} to
\begin{eqnarray}
\tilde{H}\tilde{\psi}_{n,\ell}=0,
\end{eqnarray}
where the transformed Hamilton operator is given by
\begin{subequations}
\begin{equation}
  \tilde{H} = G^{-1} \cdot H \cdot G,
\end{equation}
and the transformed wave function is written as
\begin{equation}\label{43b}
 \psi_{n,\ell}= G \cdot \tilde{\psi}_{n,\ell} .
\end{equation}
\end{subequations}
In Eq. \eqref{43b}, the function $\tilde{\psi}_{n,\ell}(r)$ is
\begin{equation}
\tilde{\psi}_{n,\ell}(r)=\sum_{k=0}^{n}a_{k}r^{k},\qquad n=1,2,3,  \dots
\end{equation}
All QES differential equations transform under the sl(2) Lie algebra with the generators
\begin{equation}\label{equ22}
\begin{split}
&J_{n}^{+}=r^{2}\frac{d}{dr}-nr,\\
&J_{n}^{0}=r\frac{d}{dr}-\frac{n}{2},\\
&J_{n}^{-}=\frac{d}{dr},\\
\end{split}
\end{equation}
which satisfy the sl(2) commutation relations:
\begin{equation}
\begin{split}
&[J^{\mp},J^{0}]=\pm J^{\pm},\\
&[J^{+},J^{-}]=-2J^{0}.
\end{split}
\end{equation}
Therefore, the operator $\tilde{H}$ can be represented as a quadratic combination of the sl(2) generators:
\begin{eqnarray}
\tilde{H} &=& C_{++} J_n^+ J_n^+ +C_{+0} J_n^+ J_n^0 + C_{+-} J_n^+ J_n^- + C_{0-} J_n^0 J_n^- + C_{--} J_n^- J_n^- + C_{+} J_n^+ + C_{0} J_n^0 + C_{-} J_n^- +C.
\end{eqnarray}
In general, any one-dimensional QES Hamiltonian can be transformed into
\begin{equation}\label{48}
\tilde H=P_{4}\frac{d^{2}}{dr^{2}}+P_{3}\frac{d}{dr}+P_{2},
\end{equation}
where $\{P_{4,3,2}\}$ are the polynomials
\begin{equation}\label{49}
\begin{split}
&{P_{4}} = {C_{ +  + }}{r^4} +{C_{ +  0 }}{r^3}+ {C_{ +  - }}{r^2} + {C_{0 - }}r + {C_{ -  - }},\\
&{P_{3}} = {C_{ +  + }}(2 - 2n){r^3} + ({C_ + } + {C_{ + 0}}(1 - \frac{{3n}}{2})){r^2}+ ({C_0} - n{C_{ +  - }})r + ({C_ - } - \frac{n}{2}{C_{0 - }}),\\
&{P_{2}} = {C_{ +  + }}n(n - 1){r^2} + (\frac{{{n^2}}}{2}{C_{ + 0}} - n{C_ + })r + (C - \frac{n}{2}{C_0}).
\end{split}
\end{equation}
Thus, we can calculate $G(r)$ in Eq. \eqref{43b} from the following relation
\begin{equation}\label{31n}
G(z)=e^{-A(r)},
\end{equation}
 one can always reduce the spectral problem for the Lie algebraic operator with the potential \cite{18}
\begin{eqnarray}
V(r)=(A')^2- A''+ P_{2}(r).
\end{eqnarray}
Sometimes, A is called prepotential. Here
\begin{eqnarray}\label{32n}
A=\int {\frac{{{P_{3}}}}{{{P_{4}}}}} dr - \log (N') \,\, , \,\, N =\pm \int {\frac{{dr}}{{\sqrt {{P_{4}}} }}}.
\end{eqnarray}
We note that, by using the QES method one can only determine a finite number of eigenvalues and their corresponding eigenfunctions algebraically \cite{Panahi}. \color{black} Here, we propose the following transformation
\begin{eqnarray}\label{25+2}
\psi_{n,\ell}(r)=r^{A}e^{-B r^{2}-Dr}\tilde{\psi}_{n,\ell}(r).
\end{eqnarray}
After straightforward calculation, we obtain the transformed Hamiltonian as
\begin{eqnarray}\label{25+3}
\tilde{H}=\frac{d^{2}}{dr^{2}}+\bigg(\frac{2A+1}{r}-4Br-2D\bigg)\frac{d}{dr}+\frac{(\lambda_{2}-2AD-D)}{r}+\lambda_{3}-4B-4AB+D^{2}
\end{eqnarray}
where  $A$, $B$ and $D$ are
\begin{subequations}\label{25+4}
\begin{eqnarray}
A&=&\sqrt{s_{1}^{2}-v_{1}^{2}+m^{2}\Omega^{2}d^{2}+\frac{(\ell-\Phi-k\chi)^{2}}{\alpha^{2}}}\ge 0, \\
B&=&\frac{1}{2}\sqrt{s_{0}^{2}-v_{0}^{2}+m^{2}\Omega^{2}b^{2}+m^{2}\omega^{2}_{c}}>0,\\
D&=&\frac{ms_{0}+E_{n,\ell}v_{0}}{\sqrt{s_{0}^{2}-v_{0}^{2}+m^{2}\Omega^{2}b^{2}+m^{2}\omega^{2}_{c}}}.
\end{eqnarray}
\end{subequations}
Then, with compare Eqs. \eqref{25+3}, \eqref{48} and \eqref{49} we find
\begin{align}\label{34+1}
&C_{++}=0, \quad \quad C_{+-}=0, \quad \quad C_{--}=0,  \quad \quad C_{-}=1-2A+\frac{n}{2}, \quad \quad  C=-nD + \lambda_{2}-2AD-D,       \nonumber\\
&C_{+0}=0, \quad \quad C_{0-}=1, \quad \quad C_{+}=-4B, \quad \quad C_{0}=-2D,              \quad \quad  -n C_{+}=\lambda_{3}-4B-4AB+D^{2}.
\end{align}
We observe a relation among the coefficients
\begin{align}\label{4z}
4B(n+1+A)=\lambda_{3}+D^{2}
\end{align}
which yields to
\begin{eqnarray}\label{26}
E^{2}_{n,\ell}&=&m^{2}+k^{2}+2m\Omega b+2m^{2}\Omega^{2}bd+2(s_{0}s_{1}-v_{0}v_{1})+\frac{2m\omega_{c}}{\alpha}(\ell-\Phi-k\chi)-\frac{(m s_{0}+E_{n,\ell}v_{0})^{2}}{s^{2}_{0}-v^{2}_{0}+m^{2}\Omega^{2}b^{2}+m^{2}\omega^{2}_{c}}\nonumber\\
&&+2\sqrt{s^{2}_{0}-v^{2}_{0}+m^{2}\Omega^{2}b^{2}+m^{2}\omega^{2}_{c}}\bigg(n+1+\sqrt{s^{2}_{1}-v^{2}_{1}+m^{2}\Omega^{2}d^{2}+\frac{(\ell-\Phi-k\chi)^{2}}{\alpha^{2}}}\bigg).
\end{eqnarray}\color{black}
We investigate the effect of the angular parameter, magnetic flux and the torsion parameters on the energy spectrum. We assume $k=b=d=\omega=1$, $s_0=0.5$, $v_0=1$, $s_1=0.1$, $v_1=1$,   and $m=2$. We  plot the energy eigenfunction versus $\alpha$, $\Phi$, $\chi$ parameters in Figs. \ref{fig:BCLCornalpha},  \ref{fig:BCLCornphi}, and \ref{fig:BCLCornchi}, respectively.

\subsection{Minimal coupling limit}
In this limit we find that the energy eigenvalue equation reduces to
 \begin{eqnarray}\label{26MCL}
E^{2}_{n,\ell}&=&m^{2}+k^{2}+2(s_{0}s_{1}-v_{0}v_{1})+\frac{2m\omega_{c}}{\alpha}(\ell-\Phi-k\chi)-\frac{(m s_{0}+E_{n,\ell}v_{0})^{2}}{s^{2}_{0}-v^{2}_{0}+m^{2}\omega^{2}_{c}}\nonumber\\
&&+2\sqrt{s^{2}_{0}-v^{2}_{0}+m^{2}\omega^{2}_{c}}\bigg(n+1+\sqrt{s^{2}_{1}-v^{2}_{1}+\frac{(\ell-\Phi-k\chi)^{2}}{\alpha^{2}}}\bigg)
\end{eqnarray}\color{black}
We note that in the case of zero external magnetic field limit we get
 \begin{eqnarray}\label{26MCLZM}
E^{2}_{n,\ell}=m^{2}+k^{2}+2(s_{0}s_{1}-v_{0}v_{1})-\frac{(m s_{0}+E_{n,\ell}v_{0})^{2}}{s^{2}_{0}-v^{2}_{0}}+2\sqrt{s^{2}_{0}-v^{2}_{0}}\bigg(n+1+\sqrt{s^{2}_{1}-v^{2}_{1}+\frac{(\ell-\Phi-k\chi)^{2}}{\alpha^{2}}}\bigg) \,\,\,\,\,
\end{eqnarray}\color{black}
\subsection{Zero magnetic field limit}
In this limit, first we examine the zero external magnetic field case. We find that Eq. \eqref{26} reduces to
\begin{eqnarray}\label{26ZMF}
E^{2}_{n,\ell}&=&m^{2}+k^{2}+2m\Omega b+2m^{2}\Omega^{2}bd+2(s_{0}s_{1}-v_{0}v_{1})-\frac{(m s_{0}+E_{n,\ell}v_{0})^{2}}{s^{2}_{0}-v^{2}_{0}+m^{2}\Omega^{2}b^{2}}\nonumber\\
&&+2\sqrt{s^{2}_{0}-v^{2}_{0}+m^{2}\Omega^{2}b^{2}}\bigg(n+1+\sqrt{s^{2}_{1}-v^{2}_{1}+m^{2}\Omega^{2}d^{2}+\frac{(\ell-\Phi-k\chi)^{2}}{\alpha^{2}}}\bigg)
\end{eqnarray}\color{black}
Then, we assume  the case where the magnetic flux does not exist. We get
\begin{eqnarray}\label{26ZMF2}
E^{2}_{n,\ell}&=&m^{2}+k^{2}+2m\Omega b+2m^{2}\Omega^{2}bd+2(s_{0}s_{1}-v_{0}v_{1})-\frac{(m s_{0}+E_{n,\ell}v_{0})^{2}}{s^{2}_{0}-v^{2}_{0}+m^{2}\Omega^{2}b^{2}}\nonumber\\
&&+2\sqrt{s^{2}_{0}-v^{2}_{0}+m^{2}\Omega^{2}b^{2}}\Bigg(n+1+\sqrt{ s^{2}_{1}-v^{2}_{1}+m^{2}\Omega^{2}d^{2}+\frac{(\ell-k\chi)^{2}}{\alpha^{2}}}\Bigg)
\end{eqnarray}\color{black}
\subsection{Linear-type potential energy limit}
We choose $v_1=s_1=0$ to examine the linear-type potential energy limit. We find that Eq. \eqref{26} reduces to
\begin{eqnarray}\label{26LTL1}
E^{2}_{n,\ell}&=&m^{2}+k^{2}+2m\Omega b+2m^{2}\Omega^{2}bd+\frac{2m\omega_{c}}{\alpha}(\ell-\Phi-k\chi)-\frac{(m s_{0}+E_{n,\ell}v_{0})^{2}}{s^{2}_{0}-v^{2}_{0}+m^{2}\Omega^{2}b^{2}+m^{2}\omega^{2}_{c}}\nonumber\\
&&+2\sqrt{s^{2}_{0}-v^{2}_{0}+m^{2}\Omega^{2}b^{2}+m^{2}\omega^{2}_{c}}\bigg(n+1+\sqrt{m^{2}\Omega^{2}d^{2}+\frac{(\ell-\Phi-k\chi)^{2}}{\alpha^{2}}}\bigg)
\end{eqnarray}\color{black}
Then, we consider the minimal coupling limit. We get
\begin{eqnarray}\label{26LTL2}
E^{2}_{n,\ell}&=&m^{2}+k^{2}+\frac{2m\omega_{c}}{\alpha}(\ell-\Phi-k\chi)-\frac{(m s_{0}+E_{n,\ell}v_{0})^{2}}{s^{2}_{0}-v^{2}_{0}+m^{2}\omega^{2}_{c}}+2\sqrt{s^{2}_{0}-v^{2}_{0}+m^{2}\omega^{2}_{c}}\bigg(n+1+\frac{\big|\ell-\Phi-k\chi \big|}{\alpha}\bigg). \,\,\,\,\,\,\,\,\,\,
\end{eqnarray}\color{black}
If we assume that the vector and scalar potential energies have the same magnitude, we find
\begin{eqnarray}\label{26LTL3}
E^{2}_{n,\ell}&=&m^{2}+k^{2}+\frac{2m\omega_{c}}{\alpha}(\ell-\Phi-k\chi)-\frac{s_{0}^2(E_{n,\ell}+m)^{2}}{m^{2}\omega^{2}_{c}}+2m\omega_{c}\bigg(n+1+\frac{\big|\ell-\Phi-k\chi \big|}{\alpha}\bigg). \,\,\,\,\,\,\,\,\,\,
\end{eqnarray}\color{black}
\subsection{Coulomb-type potential energy limit}
We choose $v_0=s_0=0$ to explore the Coulomb-type potential energy limit. We find that Eq. \eqref{26} deduces
\begin{eqnarray}\label{26CTL1}
E^{2}_{n,\ell}&=&m^{2}+k^{2}+2m\Omega b+2m^{2}\Omega^{2}bd+\frac{2m\omega_{c}}{\alpha}(\ell-\Phi-k\chi)\nonumber \\
&&+2\sqrt{m^{2}\Omega^{2}b^{2}+m^{2}\omega^{2}_{c}}\bigg(n+1+\sqrt{s^{2}_{1}-v^{2}_{1}+m^{2}\Omega^{2}d^{2}+\frac{(\ell-\Phi-k\chi)^{2}}{\alpha^{2}}}\bigg)
\end{eqnarray}\color{black}
In the $\Omega=0$ limit, we get
\begin{eqnarray}\label{26CTL2}
E^{2}_{n,\ell}&=&m^{2}+k^{2}+\frac{2m\omega_{c}}{\alpha}(\ell-\Phi-k\chi)+2 m\omega_{c} \bigg(n+1+\sqrt{s^{2}_{1}-v^{2}_{1}+\frac{(\ell-\Phi-k\chi)^{2}}{\alpha^{2}}}\bigg)
\end{eqnarray}\color{black}
If we assume that the vector and scalar potential energies have the same magnitude value, we find
\begin{eqnarray}\label{26CTL3}
E^{2}_{n,\ell}&=&m^{2}+k^{2}+\frac{2m\omega_{c}}{\alpha}(\ell-\Phi-k\chi)+2 m\omega_{c} \bigg(n+1+\frac{\big| \ell-\Phi-k\chi \big|}{\alpha}\bigg)
\end{eqnarray}\color{black}
\subsection{A case of a mixture of linear and Coulomb-type potential energies}
We find it interesting to examine the energy eigenvalue function in the limits where the scalar and vector potential energies are  Coulomb-type and linear-type, respectively. Therefore, we choose $s_0=v_1=0$ values. We find
\begin{eqnarray}\label{26Mmix1}
E^{2}_{n,\ell}&=&m^{2}+k^{2}+2m\Omega b+2m^{2}\Omega^{2}bd+\frac{2m\omega_{c}}{\alpha}(\ell-\Phi-k\chi)-\frac{ v_{0}^2 E_{n,\ell}^{2}}{m^{2}\Omega^{2}b^{2}+m^{2}\omega^{2}_{c}-v^{2}_{0}}\nonumber\\
&&+2\sqrt{m^{2}\Omega^{2}b^{2}+m^{2}\omega^{2}_{c}-v^{2}_{0}}\bigg(n+1+\sqrt{s^{2}_{1}+m^{2}\Omega^{2}d^{2}+\frac{(\ell-\Phi-k\chi)^{2}}{\alpha^{2}}}\bigg)
\end{eqnarray}\color{black}
Alike, we choose $s_1=v_1=0$ values to explore the scalar and vector potential energies  that are linear-type  and Coulomb-type, respectively. We find
\begin{eqnarray}\label{26mix2}
E^{2}_{n,\ell}&=&m^{2}+k^{2}+2m\Omega b+2m^{2}\Omega^{2}bd+\frac{2m\omega_{c}}{\alpha}(\ell-\Phi-k\chi)-\frac{s_{0}^2m^2 }{s^{2}_{0}+m^{2}\Omega^{2}b^{2}+m^{2}\omega^{2}_{c}}\nonumber\\
&&+2\sqrt{s^{2}_{0}+m^{2}\Omega^{2}b^{2}+m^{2}\omega^{2}_{c}}\bigg(n+1+\sqrt{m^{2}\Omega^{2}d^{2}+\frac{(\ell-\Phi-k\chi)^{2}}{\alpha^{2}}-v^{2}_{1}}\bigg)
\end{eqnarray}\color{black}
In the minimal length coupling limit, these equations reduce to
\begin{subequations}\label{26Mmix1ab}
  \begin{eqnarray}
   E^{2}_{n,\ell} &=& m^{2}+k^{2}+\frac{2m\omega_{c}}{\alpha}(\ell-\Phi-k\chi)-\frac{ v_{0}^2 E_{n,\ell}^{2}}{m^{2}\omega^{2}_{c}-v^{2}_{0}}+2\sqrt{m^{2}\omega^{2}_{c}-v^{2}_{0}}\bigg(n+1+\sqrt{s^{2}_{1}+\frac{(\ell-\Phi-k\chi)^{2}}{\alpha^{2}}}\bigg),\,\,\,
   \,\,\,\,\,\,\,\, \\
    E^{2}_{n,\ell} &=& m^{2}+k^{2}+\frac{2m\omega_{c}}{\alpha}(\ell-\Phi-k\chi)-\frac{s_{0}^2m^2 }{s^{2}_{0}+m^{2}\omega^{2}_{c}}+2\sqrt{s^{2}_{0}+m^{2}\omega^{2}_{c}}\bigg(n+1+\sqrt{\frac{(\ell-\Phi-k\chi)^{2}}{\alpha^{2}}-v^{2}_{1}}\bigg).
  \end{eqnarray}
\end{subequations}
\subsection{AB effect and degeneracy}
We see that in $\Phi_{B}\neq0, \chi \neq0$,  or $\Phi_{B}\neq 0$,$\chi = 0$ cases the energy eigenvalues depend on the magnetic flux. Therefore, a change of $\Phi_B$ should alter the eigenvalues. However, energy eigenvalue can remain at the same value if the magnetic quantum number is modified.

We observe degeneracy in energy states. For example, in the case of $\Phi_{B}=\chi = 0$ with $d=\omega_c=0$ and $s_1=v_1$, the following states are degenerate
\begin{itemize}
  \item For $\alpha=1$,
        \begin{subequations}
        \begin{eqnarray}
         E_{4,    0} =  E_{3,\mp 1} =  E_{2,\mp 2} &=& E_{1,\mp 3}, \\
                        E_{3,    0} =  E_{2,\mp 1} &=& E_{1,\mp 2}, \\
                                       E_{2, 0   } &=& E_{1,\mp 1}.
         \end{eqnarray}
         \end{subequations}
  \item For $\alpha=\frac{1}{2}$,
   \begin{subequations}
   \begin{eqnarray}
    E_{7,    0} =  E_{5,\mp 1} =  E_{3,\mp 2} &=& E_{1,\mp 3}, \\
                   E_{5,    0} =  E_{3,\mp 1} &=& E_{1,\mp 2}, \\
                                  E_{3,    0} &=& E_{1,\mp 1}.
  \end{eqnarray}
  \end{subequations}
\end{itemize}

\subsection{Wave function solution}
\begin{figure}[ht!]
	\centering
	\includegraphics[width=0.8\textwidth]{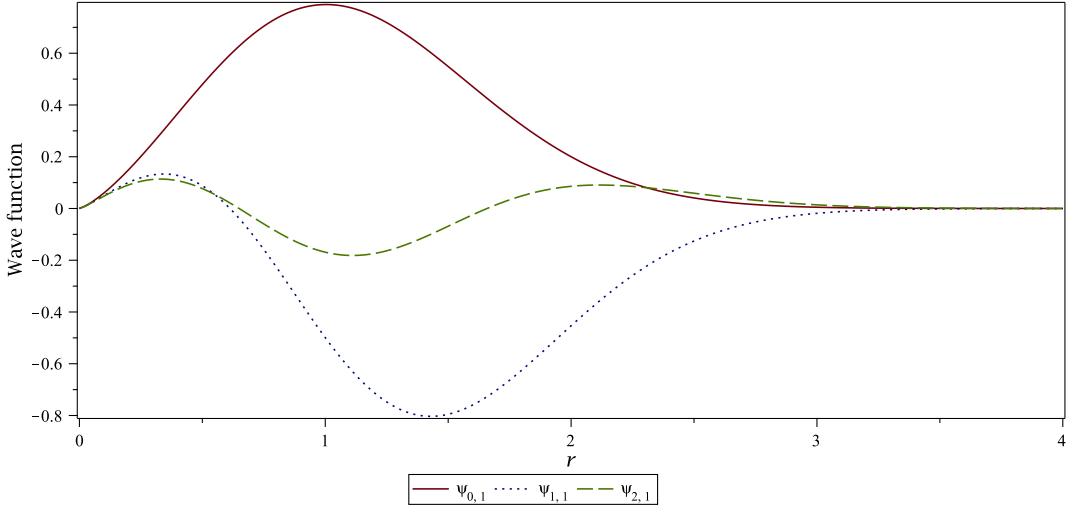}
	\caption{Unnormalized wave functions versus the distance}
	\label{fig:wf2}
\end{figure}

{In order to obtain the wave function, we use the recursion relation}
\begin{eqnarray}\label{40+1n}
a_{j+2}=\frac{2D(j+1)-\lambda_{2}+2AD+D}{(j+2)(2A+2+j)}a_{j+1}+\frac{8Bj-\lambda_{3}+4AB-D^{2}}{(j+2)(2A+2+j)}a_{j}.
\end{eqnarray}
For the first excited states, we employ the boundary condition $a_{-1}=0$. Then, the recursion relation yields to
\begin{eqnarray}
a_{1}=\frac{2AD-\lambda_{2}+D}{2A+1}a_{0}.
  \end{eqnarray}
Therefore, we express the unnormalized wave-function of the first excited state as
\begin{eqnarray}\label{42}
\psi_{1,\ell}(r)=r^{A}e^{-Br^{2}-Dr}\left(1+\frac{2AD-\lambda_{2}+D}{2A+1}r\right).
\end{eqnarray}
where  $a_{0}=1$. Alike, for the second excited-state, we find
\begin{eqnarray}
a_{2}=\frac{3D-\lambda_{2}+2AD}{4(A+1)}a_{1}+\frac{-\lambda_{3}+4AB-D^{2}}{4(A+1)}a_{0}
\end{eqnarray}
Then, we obtain the  unnormalized wave function of the  second excited state as
\begin{eqnarray}
\psi_{2,\ell}(r)=r^{A}e^{-Br^{2}-Dr}(1+a_{1}r+a_{2}r^{2}).
\end{eqnarray}
where $a_{0}=1$. We take $s_0=s_1=v_0=v_1=b=d=k=\chi=\omega_c=\Phi=1$, $\alpha=0.8$, $\Omega=0.2$, $m=2$ and evaluate some of the energy eigenvalues to demonstrate the corresponding wave functions. We present our result in Fig. \ref{fig:wf2}.

\section{Conclusion}\label{sec5}
In this manuscript, we considered a generalized K-G oscillator within an external uniform magnetic field in addition to an internal magnetic flux in a space-time with a space-time dislocation. In a non-minimal coupling, we explored solutions for a charged relativistic particle under the pseudo harmonic and Cornell-type potential energies by employing either the NU or QES method.  We observed that the NU method is reliable for the pseudo harmonic potential energy, while for the Cornell-type potential the QES method. For both potential energies, we derived an energy spectrum function, and demonstrated the dependence of the energy spectrum to the angular, magnetic flux and torsion parameters. Then, we analyzed the obtained results in various limits such as the minimal coupling, zero external magnetic fields. We discussed the AB effect on the energy spectrum.  We observed that the magnetic flux and torsion have a role in the degeneracy of the eigenstates. We plotted the ground and first two excited state's wavefunctions to demonstrate the correctness of the solutions.

\section*{Acknowledgment}
{The authors thank the referee for a thorough reading of our manuscript and for constructive suggestion.} This work is supported by the Internal Project, $[2020/2209]$, of Excellent Research of the Faculty of Science of University Hradec Kr\'{a}lov\'{e}. One of the author, B.C. L\" utf\"uo\u{g}lu, was partially supported by the Turkish Science and Research Council (T\"{U}B\.{I}TAK).


\end{document}